\documentclass[
    12pt,
    reprint,
    superscriptaddress,
    amsmath,
    amssymb,
    aps,
    prb,
    floatfix
]{revtex4-2}

\usepackage{graphicx}
\usepackage{dcolumn}
\usepackage{bm}
\usepackage{braket}
\usepackage{xcolor}
\usepackage{soul}
\usepackage[normalem]{ulem}

\usepackage{float}

\makeatletter
\let\newfloat\newfloat@ltx
\makeatother

\usepackage{algorithm}
\usepackage{algpseudocode}

\usepackage[
    breaklinks=true,
    colorlinks=true,
    citecolor=blue,
    linkcolor=blue,
    urlcolor=blue
]{hyperref}

\usepackage{orcidlink}

\newcommand{\eqlabel}[1]{Eq.~\eqref{#1}}

\newcommand{\figlabel}[1]{Fig.~\ref{#1}}

\newcommand{\orcidShubham}{\orcidlink{0000-0003-2891-1669}}
\newcommand{\orcidBhargava}{\orcidlink{0009-0006-7595-8776}}
\newcommand{\orcidAnneMaria}{\orcidlink{0000-0002-4167-7769}}
\newcommand{\orcidPaolo}{\orcidlink{0000-0003-4626-2869}}
\newcommand{\orcidEnrique}{\orcidlink{0000-0002-8602-1181}}
\newcommand{\orcidNarendra}{\orcidlink{0000-0002-9673-2833}}
\newcommand{\orcidJose}{\orcidlink{0000-0001-9378-0285}}
\newcommand{\orcidYolanda}{\orcidlink{0000-0002-3744-5893}}

\begin{document}

\title{Measurement-free Preparation of Surface-Code States with Digital-Analog Counterdiabatic Drivings}

\author{Shubham~Kumar\,\orcidShubham}
\email{shubhamkumar.kumar@gmail.com}
\affiliation{Intelligent Data Analysis Laboratory (IDAL), Departament d'Enginyeria Electrònica, ETSE-UV, Universitat de València, Universitat s/n, 46100 Burjassot, Valencia, Spain}
\affiliation{Kipu Quantum, Greifswalderstrasse 212, 10405 Berlin, Germany}

\author{Balaganchi~A.~Bhargava\,\orcidBhargava}
\affiliation{Kipu Quantum, Greifswalderstrasse 212, 10405 Berlin, Germany}

\author{Paolo~A.~Erdman\,\orcidPaolo}
\affiliation{Kipu Quantum, Greifswalderstrasse 212, 10405 Berlin, Germany}

\author{Anne-Maria~Visuri\,\orcidAnneMaria}
\affiliation{Kipu Quantum, Greifswalderstrasse 212, 10405 Berlin, Germany}

\author{José~D.~Martín-Guerrero,\orcidJose}
\affiliation{Intelligent Data Analysis Laboratory (IDAL), Departament d'Enginyeria Electrònica, ETSE-UV, Universitat de València, Universitat s/n, 46100 Burjassot, Valencia, Spain}

\author{Yolanda~Vives-Gilabert,\orcidYolanda}
\affiliation{Intelligent Data Analysis Laboratory (IDAL), Departament d'Enginyeria Electrònica, ETSE-UV, Universitat de València, Universitat s/n, 46100 Burjassot, Valencia, Spain}

\author{Enrique~Solano\,\orcidEnrique}
\affiliation{Kipu Quantum, Greifswalderstrasse 212, 10405 Berlin, Germany}

\author{Narendra~N.~Hegade\,\orcidNarendra}
\affiliation{Kipu Quantum, Greifswalderstrasse 212, 10405 Berlin, Germany}
\affiliation{Intelligent Data Analysis Laboratory (IDAL), Departament d'Enginyeria Electrònica, ETSE-UV, Universitat de València, Universitat s/n, 46100 Burjassot, Valencia, Spain}

\begin{abstract}
Preparing stabilizer-code states with shallow circuits is an important primitive for near-term quantum error correction in superconducting circuits, where logical-state initialization requires four-body stabilizer correlations from native one- and two-body controls. We introduce a digital-analog method for the preparation of surface-code ground-state manifolds. The method exploits a structural property of the adiabatic gauge potential, mapping stabilizer-local counterdiabatic terms to digital-analog blocks on the superconducting layout using fixed-angle two-qubit $XY$ gates, single-qubit rotations, and analog exchange-interaction evolutions. The dressed blocks generate higher-order operators used as the variational ansatz to determine the counterdiabatic terms within the Sels-Polkovnikov adiabatic-gauge-potential framework. We benchmark checkerboard plaquette-stabilizer grids up to $4\times4$, the square-grid instances correspond to rotated surface codes, while the rectangular cases probe the scaling of the method. For all grids, the proposed method generates the dominant four-body counterdiabatic basis and, at short evolution times substantially improves the ground-state preparation as compared with bare adiabatic evolution. We further generalize the construction to $n\times m$ stabilizer lattices and show that digital-analog synthesis can reduce the entangling depth by an order of magnitude. Together, these results establish a direct connection between stabilizer geometry, the structure of the counterdiabatic gauge potential, and digital-analog control for measurement-free preparation of stabilizer-code states on near-term superconducting architecture.
\end{abstract}


\maketitle

\section{Introduction}
\label{sec:introduction}

The surface code is a leading architecture for fault-tolerant quantum computation because it combines geometrically local stabilizer checks, a two-dimensional layout, and a high error threshold~\cite{Kitaev2003,Dennis2002,Fowler2012}. In superconducting processors, stabilizer measurements are typically implemented using one- and two-qubit gates on a two-dimensional connectivity graph~\cite{Arute2019,Gong.2023,Pedram.2024}. Before repeated error-correction cycles can protect encoded quantum information, however, the processor must first be initialized in, or close to, the desired stabilizer-code state. Developing shallow and hardware-native methods for this initialization is therefore an important primitive for quantum error correction.

Code-state initialization is distinct from repeated syndrome extraction where the latter relies on measurements, ancilla reset, and classical decoding, whereas the former can be formulated as a Hamiltonian-control problem. One starts from an easily prepared product state and evolves toward a target Hamiltonian whose ground space defines the desired code state~\cite{Farhi2000,Aharonov2008}. The objective is therefore not only to measure Pauli stabilizers, but to coherently generate the multi-qubit correlations associated with their common eigenspace. Continuously tunable Hamiltonian interactions can consequently provide useful resources for code-state initialization, even when such dynamics would be undesirable during repeated error-correction cycles.

Adiabatic evolution provides a coherent and measurement-free route to topological-state preparation. Hamma and Lidar showed that a topological order can be prepared by interpolating from a spin-polarized Hamiltonian to the toric-code Hamiltonian~\cite{HammaLidar2008}. Such protocols avoid stabilizer measurements during initialization, but their runtime is limited by the closing many-body gap and the finite coherence time of the processor. At experimentally accessible runtimes, diabatic excitations reduce the final ground-state fidelity. Accelerating this measurement-free preparation while preserving its coherent character is therefore a central challenge.

One such method to address this is counterdiabatic (CD) driving that suppresses diabatic transitions by adding an auxiliary Hamiltonian constructed from the adiabatic gauge potential (AGP)~\cite{Demirplak2003,Demirplak2008,Berry2009,Chen2010,Torrontegui2013,An2016}. In interacting many-body systems, the exact gauge potential is generally nonlocal and contains high-weight Pauli operators that are difficult to realize directly. The Sels-Polkovnikov variational principle addresses this problem by optimizing an approximate gauge potential within a restricted operator manifold~\cite{Sels2017}. Related nested-commutator (NC) and variational (Var) constructions have been developed for many-body CD protocols~\cite{PhysRevLett.123.090602,PhysRevB.106.155153,Visuri2026}. 

In this work, we apply Sels-Polkovnikov´s variational CD method to the coherent, measurement-free preparation of finite surface-code-like stabilizer states by developing a hardware-native realization of the resulting CD dynamics. The central implementation challenge is that the relevant CD corrections are non-stoquastic terms with higher-body Pauli operators, whereas near-term superconducting processors typically provide native one- and two-qubit controls. To address this, we apply digital-analog synthesis, combining native analog interactions with short digital dressing layers~\cite{Lamata2018,Julen.2016,Mikel.2020,Gonzalez-Raya2021DigitalAnalog,Kumar_2025,Kumar_fermion_boson2025,Bhargava2026}. This is motivated from related work showing that collective or multi-qubit entangling operations can reduce the overhead associated with conventional two-qubit decompositions~\cite{SinghYan2025GlobalGates}.

We show that the stabilizer geometry can be used to select hardware-accessible CD operators generated from dressed nearest-neighbor XY interactions. The local dressing rule identified on a \(3\times3\) rotated surface code is extended to larger grids through the repeated plaquette structure and the connectivity of the hardware graph. The variational CD coefficients are then mapped to the amplitudes of the analog blocks, while unwanted generated components are identified as synthesis leakage.

We benchmark the protocol on finite checkerboard plaquette-stabilizer grids containing up to \(4\times4\) qubits by comparing adiabatic evolution with variational CD evolution implemented using fully digital and digital-analog circuits. The digital-analog construction closely reproduces the ground-state fidelity of the fully digital protocol while reducing the required entangling resources. Moreover, for the \(3\times3\) patch, we compare the coherent preparation protocols with a measurement-based stabilizer-preparation scheme under a hardware-motivated noise model. While the measurement-based protocol achieves the highest fidelity, the digital-analog CD implementation approaches its performance while remaining entirely coherent and measurement-free, substantially outperforming fully digital CD and bare adiabatic evolution. 

The paper is organized as follows. Section~\ref{sec:methods} discusses the methods used to study the finite stabilizer-state-preparation problem.  Section~\ref{sec:model} discusses the \(L_x\times L_y\) rotated surface code and   Section~\ref{sec:agp} derives the plaquette-local CD basis. Section~\ref{sec:daqc_method} introduces the digital-analog quantum computing method to encode the CD terms. Section~\ref{sec:surface_code_results} describes the numerical simulation results. Section \ref{sec:3x3} discusses the minimal $3\times3$ rotated surface code and the main algorithm. Section~\ref{sec:scaling} discusses the scaling of the method including entangling-depth comparison. Section~\ref{sec:noisesimulation} discusses the simulation with a noise model. Section \ref{sec:discussion_conclusion} explains the counclusions and we end the paper with an outlook in Section~\ref{sec:outlook}.

\begin{figure*}
\centering
\includegraphics[width=0.95\textwidth]{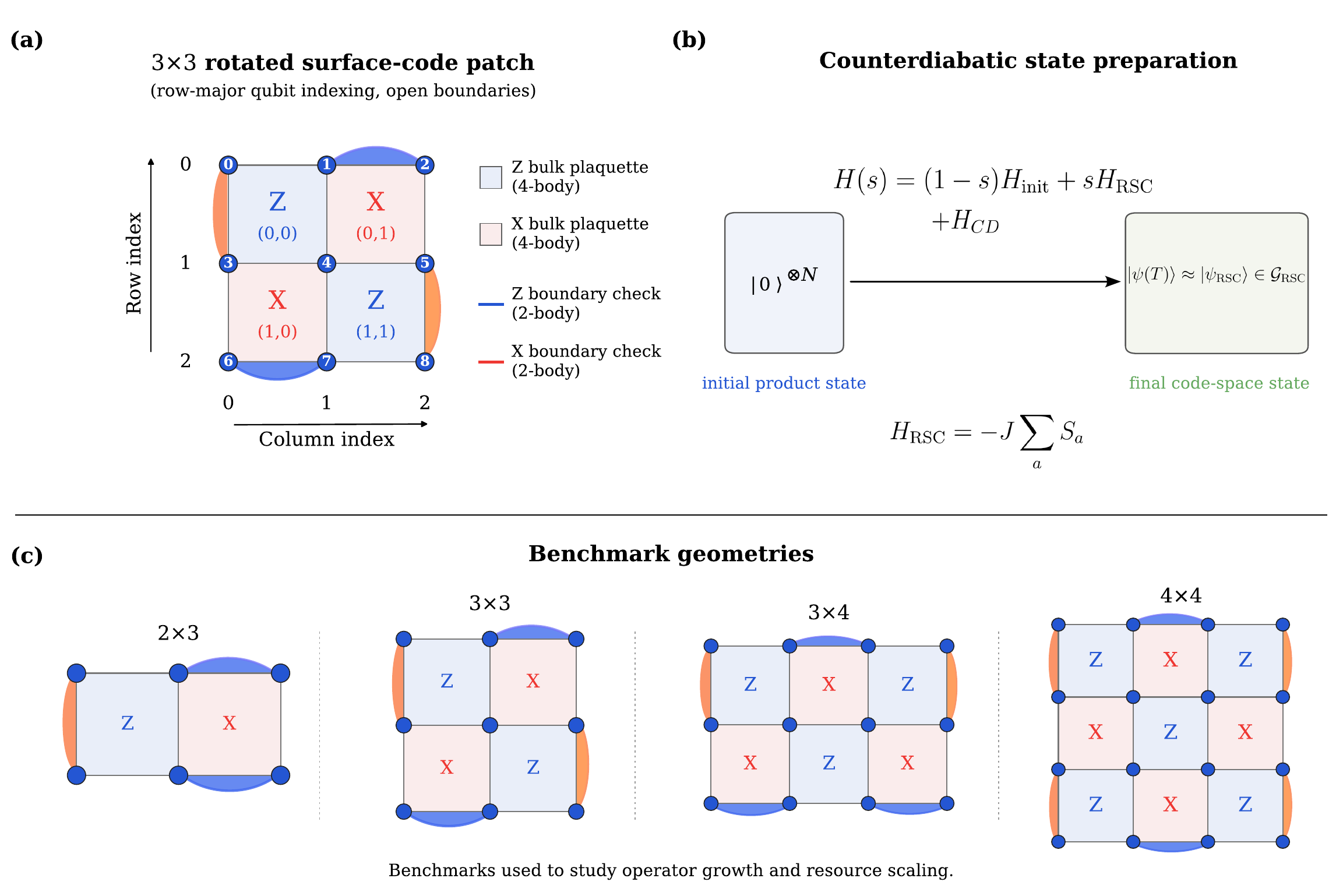}
\caption{
Surface-code-like benchmark geometry used for CD state preparation. 
(a) A \(3\times3\) qubit grid with row-major indexing. Each elementary \(2\times2\) cell defines a four-body plaquette stabilizer, with \(Z\)- and \(X\)-type plaquettes assigned in a checkerboard pattern. 
(b) The state-preparation path interpolates from the \(Z\)-polarized product state \(\ket{0}^{\otimes N}\) to the commuting plaquette-stabilizer Hamiltonian \(H_{\mathrm{RSC}}=-J\sum_p S_p\). (c) Benchmark grids used to study operator growth and resource scaling. The simulated instances are \(2\times3\), \(3\times3\), \(3\times4\), and \(4\times4\).
}
\label{fig:surface_code_benchmark_geometry}
\end{figure*}

\section{Methods}
\label{sec:methods}

\subsection{Surface-Code Ground States}

\label{sec:model}

Our target is a state in the simultaneous \(+1\) eigenspace of a set of mutually commuting stabilizers. Preparing such a state requires satisfying local stabilizer constraints while coherently generating correlations that are absent from the initial product state. Rotated surface-code Hamiltonians therefore provide a useful setting for testing whether digital-analog counterdiabatic protocols can prepare structured many-body states using hardware-native entangling interactions. We consider rectangular \(L_x\times L_y\) qubit arrays with row-major indexing
\begin{equation}
q(x,y)=xL_y+y,\hspace{1mm} x=0,\ldots,L_x-1,\quad y=0,\ldots,L_y-1,
\label{eq:row_major_indexing}
\end{equation}
so that the total number of qubits is \(N=L_xL_y\). The state-preparation path interpolates between a simple initial Hamiltonian \(H_{\mathrm{init}}\) and the rotated surface-code Hamiltonian \(H_{\mathrm{RSC}}\),
\begin{equation}
H(s)=(1-s)H_{\mathrm{init}}+sH_{\mathrm{RSC}},\qquad s\in[0,1],
\label{eq:sc_interpolation}
\end{equation}
where
\begin{equation}
H_{\mathrm{init}}=-h\sum_{i=0}^{N-1}Z_i.
\label{eq:sc_initial_hamiltonian}
\end{equation}
We set \(\hbar=1\), report energies in units of \(h\), and express time through the dimensionless variable \(ht\). The initial state is \(\ket{\psi_0}=\ket{0}^{\otimes N}\).

Each elementary \(2\times2\) cell supports a four-body bulk stabilizer, while the open boundaries support two-body stabilizers. The stabilizers are assigned as \(X\) or \(Z\) type according to the checkerboard pattern. For a stabilizer \(a\) acting on the qubit set \(Q(a)\), we define
\begin{equation}
S_a=
\begin{cases}
\displaystyle\prod_{q\in Q(a)}X_q,\\[1mm]
\displaystyle\prod_{q\in Q(a)}Z_q,
\end{cases}
\label{eq:checkerboard_stabilizers}
\end{equation}
The target Hamiltonian is
\begin{equation}
H_{\mathrm{RSC}}=-J\sum_a S_a.
\label{eq:sc_target}
\end{equation}

The benchmark geometry is shown in Fig.~\ref{fig:surface_code_benchmark_geometry}. An \(L_x\times L_y\) patch contains \((L_x-1)(L_y-1)\) four-body plaquettes together with the corresponding two-body boundary stabilizers. If \(N_S\) denotes the total number of stabilizers, their mutual commutativity implies the ground-state energy
\begin{equation}
E_0=-JN_S.
\label{eq:target_ground_energy_general}
\end{equation}

Because \(H_{\mathrm{SC}}\) has a degenerate ground space, the relevant objective is the overlap with the complete code space and we use the ground-space projector
\begin{equation}
P_{\mathrm{GS}}=\prod_{c\in\mathcal{C}}\frac{1+S_c}{2}
\label{eq:gs_projector}
\end{equation}
and define the ground-state fidelity as
\begin{equation}
F_{\mathrm{GS}}(t)=\bra{\psi(t)}P_{\mathrm{GS}}\ket{\psi(t)}.
\label{eq:gs_fidelity}
\end{equation}
We also evaluate the target energy
\begin{equation}
E(t)=\bra{\psi(t)}H_{\mathrm{SC}}\ket{\psi(t)},
\label{eq:target_energy}
\end{equation}
which approaches \(E_0\) when all stabilizer constraints are satisfied. Since the interpolation begins from a \(Z\)-polarized product state, the \(Z\)-type stabilizers are initially favored, whereas generating the correlations required by the \(X\)-type stabilizers constitutes the nontrivial part of the dynamics.

We study the finite grids upto \(4\times4\) and 16 qubits, depicted in Fig. \ref{fig:surface_code_benchmark_geometry}. These instances allow us to examine how the numbers of stabilizer constraints, CD generators, digital entangling gates, and digital-analog blocks scale with system size. State preparation and noisy dynamics are simulated for these patches, while the resource analysis is performed for all system sizes. For the numerical benchmark, we first use the \(3\times3\) patch, that is the rotated surface code with distance $d=3$. The bulk stabilizers for this system are
\begin{align}
S_{Z,1}^{(4)}&=Z_0Z_1Z_3Z_4,&S_{X,1}^{(4)}&=X_1X_2X_4X_5,\nonumber\\
S_{X,2}^{(4)}&=X_3X_4X_6X_7,&S_{Z,2}^{(4)}&=Z_4Z_5Z_7Z_8,
\end{align}
and the boundary stabilizers are
\begin{align}
S_{X,1}^{(2)}&=X_0X_3,&S_{X,2}^{(2)}&=X_5X_8,\nonumber\\
S_{Z,1}^{(2)}&=Z_1Z_2,&S_{Z,2}^{(2)}&=Z_6Z_7.
\label{eq:sc_stabilizers_3x3}
\end{align}
Thus, \(N_S=8\) and, for \(J=h\), the dimensionless ground-state energy is \(E_0/h=-8\).

\subsection{Stabilizer-Local Counterdiabatic Gauge Potential}
\label{sec:agp}

Counterdiabatic (CD) driving suppresses diabatic excitations by augmenting the interpolating Hamiltonian with an auxiliary control term,
\begin{equation}
H_{\mathrm{CD}}(t)=H(s(t))+\dot{s}(t)A_s,
\label{eq:cd_hamiltonian}
\end{equation}
where \(A_s\) is the adiabatic gauge potential (AGP). Since the exact AGP is generally highly nonlocal, we employ the variational construction of Sels and Polkovnikov~\cite{Sels2017},
\begin{equation}
A_s=\sum_\mu\alpha_\mu(s)O_\mu,
\label{eq:varcd_ansatz}
\end{equation}
where the coefficients \(\alpha_\mu(s)\) are obtained by minimizing the Hilbert-Schmidt action
\begin{equation}
\mathcal{S}[A_s]
=
\operatorname{Tr}
\!\left[
\partial_sH+i[A_s,H]
\right]^2.
\label{eq:sels_action}
\end{equation}
The coefficients are pre-computed classically for all the interpolation points and specify the CD operator at every point. The output is a set of operators with optimal coefficients. 
\begin{figure*}
\centering
\includegraphics[width=1\textwidth]{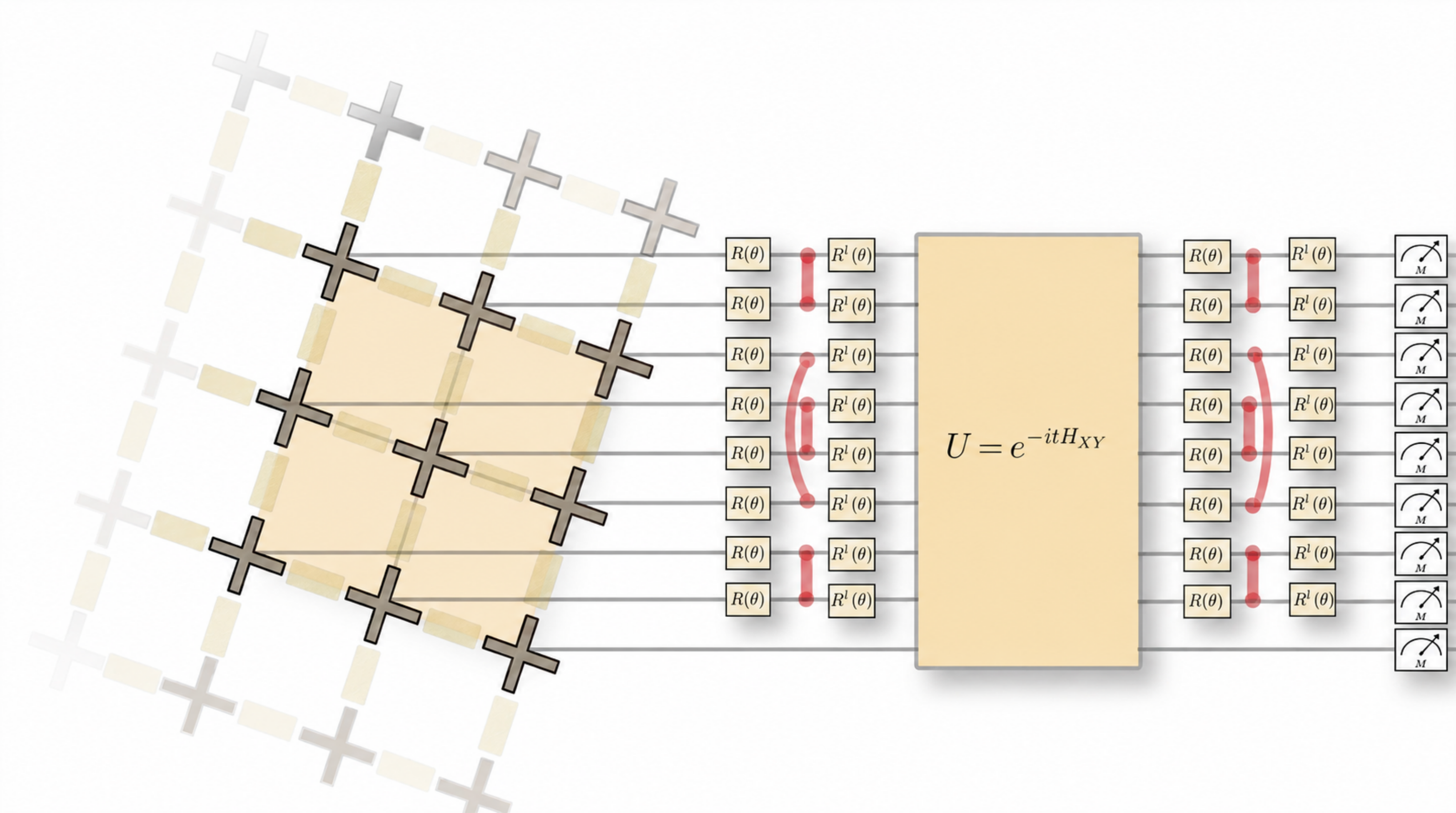}
\caption{
Digital-analog construction for higher-order Hamiltonian ansatz. A two-dimensional superconducting-qubit processor is modeled as a grid of transmon qubits connected by tunable couplers. Activating the couplers generates native \(XY\)-exchange interactions on the hardware graph. Digital dressing layers, built from fixed-angle two-qubit \(XY\)-type gates (red curves) and single-qubit rotations, are applied on a matching set consisting of disjoint edges. Conjugating the native exchange block sequentially by these matching-dependent dressing circuits generates a structured pool of multi-body Pauli operators relevant to plaquette-local CD synthesis.
}
\label{fig:daqc_model}
\end{figure*}

The remaining task is therefore to determine the operator basis \(\{O_\mu\}\). For the surface-code interpolation considered here, this basis follows directly from the stabilizer geometry. The initial Hamiltonian is a transverse \(Z\)-field, Eq. \ref{eq:sc_initial_hamiltonian}
while the target Hamiltonian consists of \(X\)- and \(Z\)-type stabilizers. Since the \(Z\)-type stabilizers commute with the driver, they do not contribute to the leading local AGP. The CD terms arise from the \(X\)-type stabilizers through
\begin{equation}
i[H_{\mathrm{init}},X_{i}X_{j}X_{k}X_{l}]
=
2
\sum_{j} \left(
Y_{j}
\prod_{k\neq j}
X_{k}\right),
\label{eq:driver_stabilizer_commutator}
\end{equation}
showing that each stabilizer generates operators obtained by replacing one \(X\) with a \(Y\). For example, a bulk plaquette $X_aX_bX_cX_d$ produces the operators
\begin{equation}
Y_aX_bX_cX_d,\,
X_aY_bX_cX_d,\,
X_aX_bY_cX_d,\,
X_aX_bX_cY_d
\label{eq:bulk_cd_basis}
\end{equation}
whereas a boundary stabilizer $X_aX_b$
gives
\begin{equation}
Y_aX_b,\,
X_aY_b.
\label{eq:boundary_cd_basis}
\end{equation}
The local VarCD basis is obtained by collecting these operators from every \(X\)-type stabilizer. For the \(3\times3\) patch, this gives twelve candidate CD operators: eight four-body terms from the two bulk plaquettes and four two-body terms from the boundary stabilizers.

For the four-body CD terms, rather than implementing individually, we aim to generate them using dressed analog $XY$ blocks. The variational coefficients $\alpha_\mu$ set the target amplitudes of the CD terms, that are eventually translated to the exchange couplings $J_{ij}$ through the DAQC synthesis described in the next section. The remaining weight-two boundary terms are implemented digitally.

As a reference, we also consider the first-order nested-commutator ansatz~\cite{PhysRevLett.123.090602},
\begin{equation}
O_1=i[H_{\mathrm{init}},H_{\mathrm{SC}}],
\label{eq:o1_generator}
\end{equation}
with
\begin{equation}
A_s^{\mathrm{NC}}=\alpha_1(s)O_1.
\label{eq:nc_cd_agp}
\end{equation}
For the present interpolation, \(O_1\) contains the same bulk and boundary single-\(Y\) operators as the variational basis, differing only by overall coefficients inherited from the commutator. Nested-commutator CD fixes the relative amplitudes of all operators through a single coefficient \(\alpha_1(s)\), whereas the variational construction determines each coefficient \(\alpha_\mu(s)\) independently. This additional variational freedom enables a more accurate approximation to the exact AGP while preserving the same local operator support.

The next section shows how the resulting four-body CD operator basis is synthesized via hardware-native digital-analog circuits.

\subsection{Digital-analog synthesis}
\label{sec:daqc_method}

We now describe how the plaquette-local CD operators are synthesized from digital-analog synthesis. The synthesis reverses the conventional workflow and proceeds in two stages, digital-analog circuits first generate a hardware-native Pauli pool, which is then used as the variational AGP ansatz in the Sels-Polkovnikov equations to select the relevant CD operators and determine their coefficients. Secondly, these coefficients are mapped to the couplings $J_{ij}$ of the analog block for the actual evolution.

We consider the layout of a superconducting processor, Fig.~\ref{fig:daqc_model}, whose qubits form the vertices of a hardware graph \(G=(V,E)\), with tunable couplers defining the available nearest-neighbor edges. In a frame rotating at
the qubit frequencies, where the free single-qubit precession is removed, the native analog interaction is \cite{Blais2021} 
\begin{equation}
    H_{\mathrm{XY}}
    =
    \sum_{\langle i,j\rangle\in E}
    J_{ij}
    \left(
    X_iX_j+Y_iY_j
    \right),
    \label{eq:xy_hamiltonian}
\end{equation}
where \(J_{ij}\) is the programmable exchange amplitude on edge \(\langle i,j\rangle\). Such \(XY\)-type interactions are natural in tunable-coupler superconducting architectures and have been used as programmable analog resources in digital-analog simulation protocols~\cite{Arute2019,Gong.2023,Pedram.2024}. Here, the hardware graph is chosen to match the rectangular qubit grid supporting the surface-code-like plaquettes.

\begin{figure*}
    \centering
    \includegraphics[width=\linewidth]{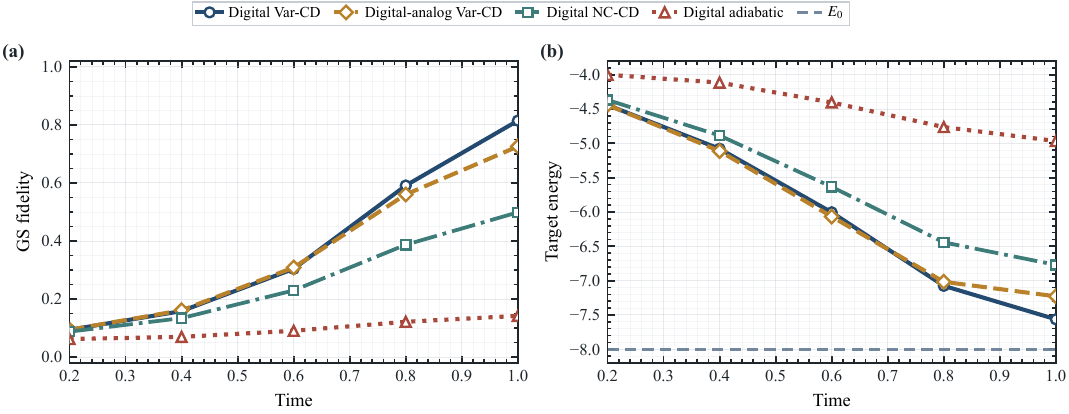}
    \caption{The ground-state fidelity and target energy, defined in Eqs.~\eqref{eq:gs_fidelity} and \eqref{eq:target_energy}, as functions of time during preparation of the ground space of \(H_{\mathrm{RSC}}\), \eqlabel{eq:sc_target} on a \(3\times3\) square patch. Time is reported in units of \(h^{-1}\). The dashed line in panel~(b) denotes the ground-state energy of \(H_{\mathrm{SC}}\). We set \(J=h\) and the analog couplings \(J_{ij}\) in \eqlabel{eq:xy_hamiltonian} are determined from the variational coefficients \(\alpha_\mu(s)\), Eq. (\ref{eq:varcd_ansatz}) - Eq. (\ref{eq:sels_action}) for each dressed analog block.
    }
    \label{fig:surface_code_9q_dynamics}
\end{figure*}

To determine the dressing circuits and an efficient synthesis, we use the concept of a matching, \(m\subset E\) of a hardware graph \(G=(V,E)\). A matching is a subset of pairwise disjoint edges. It specifies the qubit pairs on which the two-qubit dressing gates will be applied in parallel, defining the dressing circuit \(D_{m,\nu}\). For a matching \(m\subset E\) and dressing choice \(\nu\), the dressed analog block is
\begin{equation}
    \widetilde H_{\mathrm{XY}}
    =
    D_{m,\nu}^{\dagger}
    H_{\mathrm{XY}}
    D_{m,\nu} = e^{-i\widetilde H_{\mathrm{XY}}},
    \label{eq:dressed_xy_matching}
\end{equation}
where \(H_{\mathrm{XY}}\) is the exchange Hamiltonian, and \(D_{m,\nu}\) is a fixed dressing circuit consisting of fixed-angle two-qubit $XY$ gates on disjoint matched edges, supplemented by single-qubit rotations. 
The matching \(m\) determines which edges are dressed in parallel, while \(\nu\) specifies the local basis choice. Different choices of \(\nu\) rotate the dressed interaction into different Pauli operators, allowing the same native \(XY\) interaction to generate different Pauli strings.

Depending on the matching and dressing, this construction produces two-body, three-body and four-body operators. A schematic form of the operator growth produced by matching-based dressed exchange blocks can be written in a local Pauli frame as (see Supplementary Sec.~\ref{subsec:higher_order_interactions} for calculations)
\begin{multline}
\tilde{H}_{XY}^{\mathrm{even}}
=
  \sum_{\langle i,j\rangle}
  J_{ij}^{(\nu)}
  \left(
        P_i P_j
      + Q_i Q_j
  \right)
  \\
  +
  \sum_{\langle i,j,k,l\rangle}
  K_{jk}^{(\nu)}
  \left(
        P_i R_j R_k P_l
      + Q_i R_j R_k Q_l
  \right),
      \label{eq:generated_even_terms}
\end{multline}
with additional odd-length contributions of the form
\begin{equation}
\tilde{H}_{XY}^{\mathrm{odd}}
=
  \tilde{H}_{XY}^{\mathrm{even}}
  +
  \sum_{\langle i,j,k\rangle_L}
  L_{jk}^{(\nu)}
  \left(
        P_i R_j Q_k
      -
        Q_i R_j P_k
  \right).
      \label{eq:generated_odd_terms}
\end{equation}
Here \(P_i,Q_i,R_i\in\{X_i,Y_i,Z_i\}\) denote local Pauli axes determined by the dressing choice \(\nu\), and the effective coefficients \(J_{ij}^{(\nu)}\), \(K_{jk}^{(\nu)}\), and \(L_{jk}^{(\nu)}\) are determined by the source exchange couplings and the chosen basis-transform circuit. Equations~\eqref{eq:generated_even_terms} and~\eqref{eq:generated_odd_terms} are therefore not fixed target Hamiltonians. By changing the single-qubit rotations in \(D_{m,\nu}\), the same matching pattern can be read out in different local Pauli frames. In particular, suitable choices of these local frames can generate plaquette-local CD terms, Eq. (\ref{eq:bulk_cd_basis}), which form the leading plaquette-local CD basis. Details of the calculation are given in Supplementary Sec.~\ref{subsec:higher_order_interactions}. 

The variational adiabatic gauge potential is constructed directly on this hardware-generated operator pool,
\begin{equation}
A_s^{\mathrm{DAQC}}
=
\sum_{O_\mu\in\mathcal B_{\mathrm{gen}}^{\mathrm{DAQC}}}
\alpha_\mu(s)O_\mu,
\end{equation}
where $B_{\mathrm{gen}}^{\mathrm{DAQC}}$ is the ansatz pool. Solving the Sels-Polkovnikov equations determines the optimal coefficients
$\alpha_\mu(s)$
for every generated operator.
In practice, this method assigns optimal coefficients to only a small subset of the generated operators only when its commutator has a nonzero overlap with the relevant diabatic directions, while the remaining terms have negligible weight. It therefore determines both the relevant operators and their coefficients in the approximate gauge potential.

Each retained operator is traced back to the the dressing circuit \(D_{m,\nu}\) and the analog block couplings \(J_{ij}\) that generated it. These source couplings determine which analog blocks are activated in the main evolution. 

In the fast-driving \emph{impulse regime}, the CD contribution dominates the evolution~\cite{Carolan2022Impulse,GomezCadavid2024Impulse}. We compare the integrated contributions to the exponent of the propagator. For a characteristic Hamiltonian scale \(J\), the adiabatic contribution scales as \(JT\), whereas for a linear schedule \(\lambda=t/T\),
\begin{equation}
\int_0^T\dot{\lambda}(t)A_\lambda\,dt
=
\int_0^1 A_\lambda\,d\lambda,
\end{equation}
which is independent of \(T\). Thus, in the impulse regime \(JT\ll1\) (\(\hbar=1\)), the adiabatic contribution is suppressed while the integrated CD contribution remains finite. We therefore retain only the CD contribution in the DAQC implementation. This is also natural from the synthesis perspective, since the dressed \(XY\) blocks directly generate the stabilizer-local CD basis, whereas including \(H(\lambda)\) would require additional entangling operations. 

This construction is scalable because the dressing layers are matching-based. Once a local dressing pattern is found, it can be repeated across disjoint edges of a larger hardware graph. The main synthesis task is to identify the single-qubit rotations and fixed two-qubit basis transforms that map native exchange interactions into Pauli strings with the desired interaction. Using this synthesis, the coefficients of the generated higher-body operators depend only on the two-qubit couplings through $J_{ij}^s$ and do not require direct multi-qubit couplings. Meanwhile the variational selection step prevents the method from relying on an uncontrolled operator expansion.

\section{Numerical Results and Coefficient Structure}
\label{sec:surface_code_results}

\subsection{$3\times3$ Rotated Surface-Code Benchmark}
\label{sec:3x3}

We compare bare adiabatic evolution with the Var and NC-CD methods under both digital and digital-analog approaches. Algorithm~\ref{alg:varcd_daqc_surface_code} summarizes the synthesis and evolution.

We first consider the \(3\times3\) square patch to illustrate the quantum evolution and coefficient structure. It contains eight stabilizer checks with two bulk \(X\)-type and two bulk \(Z\)-type plaquettes, together with two \(X\)-type and two \(Z\)-type boundary checks. The two weight-four \(X\)-type plaquettes generate eight local single-\(Y\) CD operators, while the two weight-two \(X\)-type boundary checks generate four additional operators, giving a total of twelve CD operators. Table~\ref{tab:3x3_coeff_structure} in Supplementary Sec. \ref{tab:3x3_coeff_structure} lists these operators and compares the coefficients obtained from these methods. NC-CD has the same leading operator support for this stabilizer interpolation, but its relative weights are fixed by the commutator structure. VarCD instead optimizes each coefficient independently. The DAQC-varCD protocol uses the same optimized coefficients as digital VarCD, but lacks all the operator support as the synthesis cannot cover all the operators.

For the DAQC, the matchings and dressing operators used are summarized in Table~\ref{tab:3x3_dressing_matchings} in the supplementary. The table lists the matchings and the corresponding two-qubit dressing Hamiltonians used to generate the retained CD terms.

For the \(3\times3\) patch, the hardware-generated operator pool contains 188 Pauli strings produced by the dressed analog blocks. Solving the Sels-Polkovnikov equations assigns nonzero coefficients to only six generated operators, demonstrating that the physically relevant adiabatic gauge potential occupies only a small subspace of the generated hardware-native basis. 

The corresponding ground state preparation dynamics are shown in \figlabel{fig:surface_code_9q_dynamics}. The total runtime considered is \(T=1\), the evolution is discretized into ten steps of size \(\Delta t=0.1\), and we use a linear schedule \(s=t/T\) in \eqlabel{eq:cd_hamiltonian}. Adiabatic method does not reach the target ground space within the short runtime \(T=1\). NC-CD improves the preparation, while digital VarCD gives the highest final fidelity. The hardware-native DAQC-VarCD implementation remains close to digital VarCD, showing that the selected dressed exchange blocks capture the dominant CD structure.

\begin{algorithm}
\caption{VarCD-DAQC construction and evolution}
\label{alg:varcd_daqc_surface_code}
\begin{algorithmic}[1]

\Require Hardware graph \(G=(V,E)\), initial Hamiltonian
\(H_{\mathrm{init}}\), target Hamiltonian \(H_{\mathrm{RSC}}\),
native interaction \(H_{\mathrm{XY}}\), interpolation
points \(\{s_k\}\), and time step \(\Delta t\)

\Ensure Digital-analog counterdiabatic evolution

\State Identify the \(X\)-type stabilizers of \(H_{\mathrm{RSC}}\).

\State Generate the admissible matchings \(m\) of the hardware graph.

\For{each matching \(m\) and basis choice \(\nu\)}

    \State Construct the dressing circuit \(D_{m,\nu}\).

    \State Generate the dressed analog interaction
    \[
    \widetilde H_{\mathrm{XY}}^{(m,\nu)}
    =
    D_{m,\nu}^{\dagger}
    H_{\mathrm{XY}}
    D_{m,\nu}.
    \]

    \State Identify the generated higher-body Pauli operators.

\EndFor

\State Collect the generated operators to form
\(\mathcal B_{\mathrm{gen}}^{\mathrm{DAQC}}\).

\State Construct the variational AGP ansatz
\[
A_s^{\mathrm{DAQC}}
=
\sum_{O_\mu\in\mathcal B_{\mathrm{gen}}^{\mathrm{DAQC}}}
\alpha_\mu(s)O_\mu.
\]

\State Solve the Sels-Polkovnikov
variational equations to obtain
\(\alpha_\mu\).

\State Retain the dressed analog blocks generating the required
four-body CD operators.

\State Determine the corresponding exchange couplings
\(J_{ij}\) from the precomputed VarCD coefficients
\(\alpha_\mu\).

\State Combine the four-body analog blocks with the two-body
boundary CD terms.

\For{each Trotter step \(k\)}

    \State Apply the two-body CD terms using the precomputed
    coefficients \(\alpha_\mu\).

    \For{each retained dressed analog block \((m,\nu)\)}

        \State Apply
        \[
        D_{m,\nu}^{\dagger}
        e^{-i\Delta t H_{\mathrm{XY}}}
        D_{m,\nu}
        =
        e^{-i\Delta t
        \widetilde H_{\mathrm{XY}}^{(m,\nu)}}.
        \]

    \EndFor

\EndFor

\end{algorithmic}
\end{algorithm}

\begin{figure*}
    \centering
    \includegraphics[width=\linewidth]{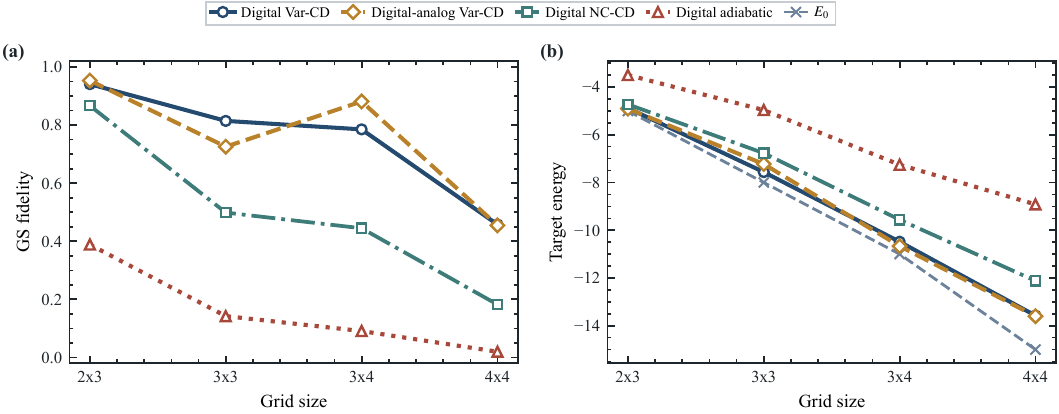}
    \caption{
    Ground-state fidelity and target energy versus grid size for the surface-code-like stabilizer benchmarks. The comparison is performed at total runtime \(T=1\) with five Trotter steps. 
    }
    \label{fig:scaling_fidelity_energy}
\end{figure*}

\subsection{Scaling with Lattice Size}
\label{sec:scaling}

We study the scalability of the construction on square lattices up to \(4\times4\). For an \(n\times n\) rotated patch with \(n^2\) data qubits, the \((n-1)^2\) bulk and \(2(n-1)\) boundary checks give \(N_{\mathrm{stab}}=n^2-1\) stabilizers. For the \(Z\)-driver convention used here, the stabilizer-local VarCD basis contains
\begin{equation}
N_{\mathrm{CD}}
=
\begin{cases}
2n(n-1), & n\ \mathrm{odd},\\
2n(n-1)-2, & n\ \mathrm{even}.
\end{cases}
\end{equation}
Both therefore scale as \(O(n^2)\).

For lattices up to \(3\times4\), the relevant DAQC blocks are identified by a classical search over admissible matchings of the hardware graph and their dressing configurations. For each candidate, the dressed native \(XY\) interaction of Eq.~\ref{eq:dressed_xy_matching} is expanded in the Pauli basis, and only configurations generating the required four-body CD operators are retained. The CD coefficients are obtained from the variational adiabatic-gauge-potential equations introduced above. 

The dominant classical cost arises from enumerating admissible matchings of the hardware graph. For matchings of fixed maximum size \(k_{\max}\), the number of candidates is bounded by \(\sum_{k=1}^{k_{\max}}\binom{|E|}{k}\), where \(|E|\) is the number of hardware edges~\cite{LovaszPlummer1986}. For a two-dimensional nearest-neighbor lattice, \(|E|=O(N)\), giving polynomial preprocessing complexity for fixed \(k_{\max}\).

For the \(4\times4\) lattice, we reuse the local matching patterns selected on the \(3\times4\) reference system and generate their valid translated, rotated, and reflected copies. This resulting pool is then used to construct the \(4\times4\) DAQC blocks, while the VarCD coefficients are recomputed for the \(4\times4\) Hamiltonian. Details of the procedure is given in the Supplementary Sec.~\ref{sec:matching_transfer}, while the results are explained below.

\begin{figure}
    \centering
    \includegraphics[width=\linewidth]{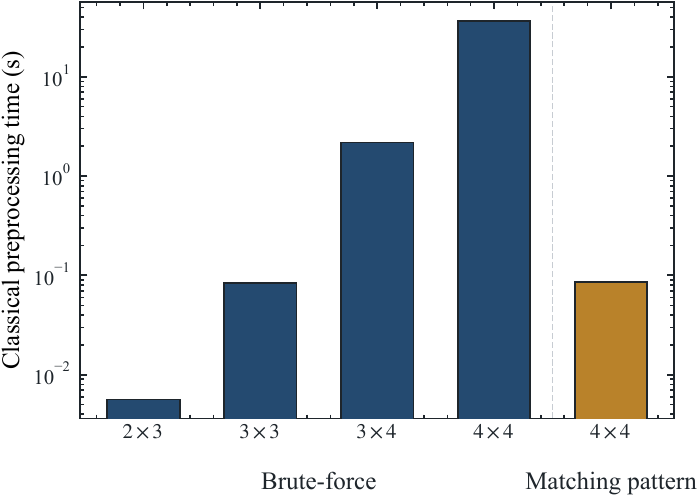}
    \caption{
    Classical preprocessing time for the matching-based DAQC construction. The first four bars use brute-force method to find the matching. The final \(4\times4\) result transfers the CD-generating matching pattern identified upto \(3\times4\) reference geometry, avoiding a new matching search.
    }
    \label{fig:preprocessing_scaling}
\end{figure}

We compare the resulting quantum-resource scaling in Fig.~\ref{fig:scaling_fidelity_energy} showing that CD protocols consistently outperform adiabatic evolution at a fixed runtime, while the DAQC results closely follow the digital VarCD results as the lattice size increases. Moreover, the \(4\times4\) result is evaluated based on the matching pattern found upto \(3\times4\) and the corresponding classical preprocessing costs shown in Fig.~\ref{fig:preprocessing_scaling}. The brute-force runtime increases rapidly from \(2\times3\) to \(4\times4\) as the number of candidate matchings grows. In contrast, the pattern-based \(4\times4\) construction evaluates only symmetry and translation embeddings of the structures identified on the \(3\times4\) lattice, substantially reducing the preprocessing cost.

\begin{figure}
    \centering
    \includegraphics[width=\linewidth]{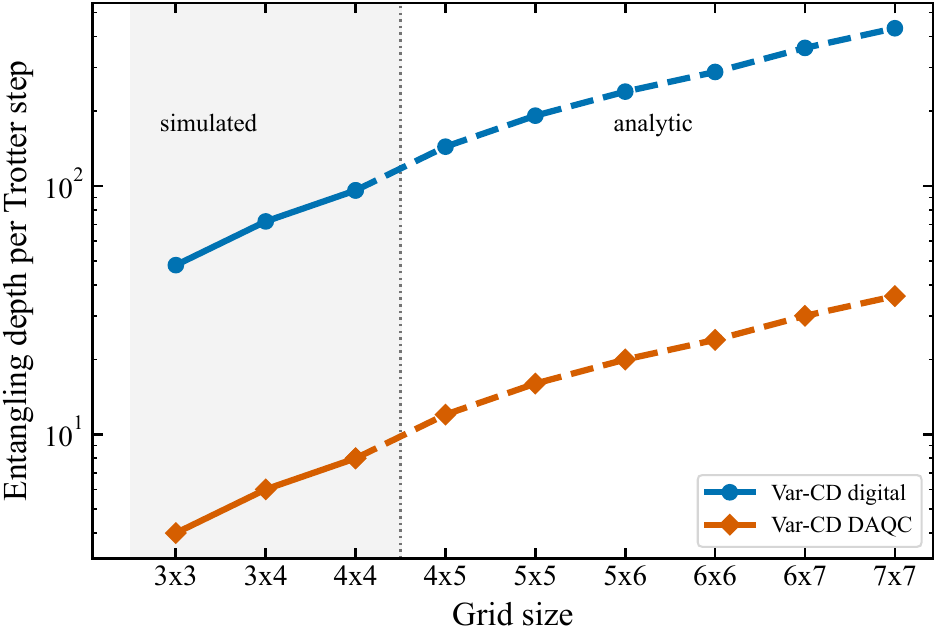}
    \caption{
    Circuit-depth scaling for the stabilizer-state preparation benchmarks at \(T=1\) for a single Trotter step. The digital curve uses two-qubit iswap gates as the entangling primitive, while the DAQC curve is based on the entangling depth of the dressed $J_{ij}(X_iX_j+Y_iY_j)$. The shaded region denotes simulated instances and the unshaded region the analytic extrapolation.
    }
    \label{fig:circuit_depth_scaling}
\end{figure}

Figure~\ref{fig:circuit_depth_scaling} compares the circuit-depth scaling of the digital and DAQC implementations. In the digital approach, each many-body Pauli evolution is decomposed and transpiled into the \(i\mathrm{SWAP}\) entangling basis, so the circuit depth increases with both the number and the weight of the Pauli terms. In the DAQC construction, multiple CD operators are generated collectively within the same dressed analog block, while the disjoint two-qubit dressing operations associated with a matching are applied in parallel. The DAQC depth is therefore governed by the number of retained matching blocks and by the entangling depth required to dress and realize each analog block, rather than by the number of individual Pauli strings.

\subsection{Trotterization Robustness}
\label{sec:trotter_hardware_times}

We discuss the robustness of different methods against trotterized evolutions. This explores the feasibility of the protocols considered for rapid state preparation. For a fixed total numerical evolution time \(T\), we divide the protocol into \(N_t\) intervals of duration
\begin{equation}
\Delta t=\frac{T}{N_t},
\label{eq:numerical_time_step}
\end{equation}
and evaluate the Hamiltonians at the midpoint
\begin{equation}
t_k=\left(k-\frac{1}{2}\right)\Delta t,
\qquad
k=1,\ldots,N_t.
\label{eq:midpoint_times}
\end{equation}

\begin{figure}
\centering
\includegraphics[width=\columnwidth]{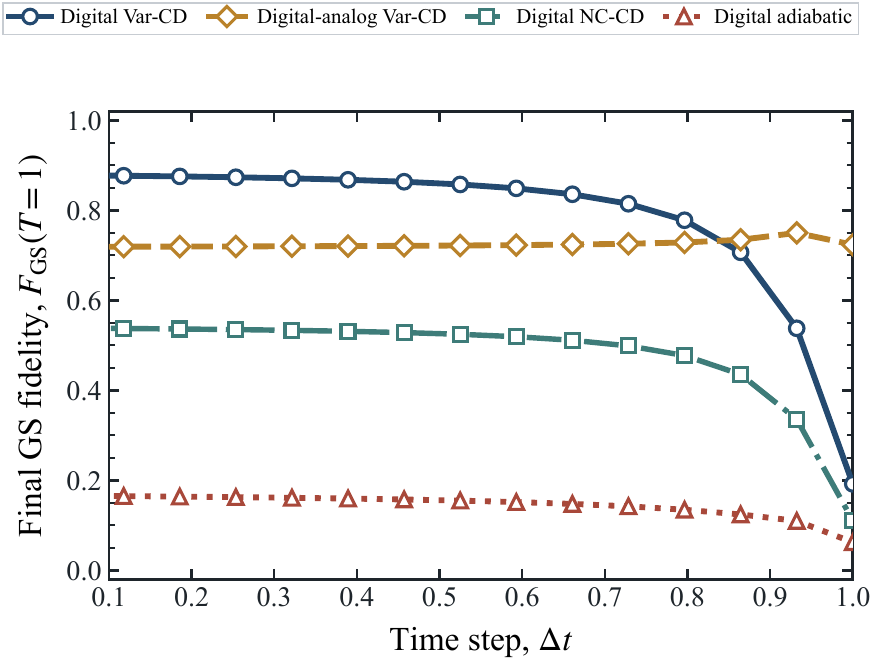}
\caption{
Final ground-space fidelity as a function of the numerical time step \(\Delta t=T/N_t\) at fixed total evolution time \(T=1\). Smaller time steps correspond to a finer discretization of the protocol. Digital VarCD displays the strongest time-step dependence, whereas the digital-analog VarCD result is comparatively insensitive to \(\Delta t\) because each dressed native interaction is implemented as a complete analog block.
}
\label{fig:timestep_convergence}
\end{figure}

Figure~\ref{fig:timestep_convergence} shows the final ground-state fidelity as a function of increasing \(\Delta t\) at fixed \(T\). The digital VarCD result displays the strongest dependence on the time-step size because its Hamiltonian contains the adiabatic terms together with several mutually noncommuting CD Pauli strings. A first-order decomposition therefore introduces a visible product-formula error. The adiabatic and NC-CD protocols also vary with \(\Delta t\), although they converge to lower final fidelities for the short total evolution time considered here.

The digital-analog VarCD (CD only) result is comparatively insensitive to the numerical time step. In this implementation, the two-body CD terms are applied digitally, whereas the four-body terms are generated by dressed analog blocks. Each analog block evolves under the complete native interaction rather than under a Pauli-by-Pauli decomposition. The DAQC result shows a weaker dependence on the time-step size because it includes only CD part and each dressed native interaction is implemented as a complete analog block rather than decomposed into individual Pauli evolutions. This behavior can be understood from the integrated CD angle. For a linear interpolation,
\begin{equation}
s(t)=\frac{t}{T},
\qquad
\dot{s}=\frac{1}{T},
\label{eq:linear_schedule_time}
\end{equation}
and the CD evolution in the \(k\)-th interval is governed by
\begin{equation}
U_{\mathrm{CD},k}
\simeq
\exp\left[
-i\dot{s}\,
A_s(s_k)\Delta t
\right]
=
\exp\left[
-iA_s(s_k)\Delta s
\right],
\label{eq:cd_schedule_increment}
\end{equation}
where
\begin{equation}
\Delta s=\dot{s}\Delta t=\frac{1}{N_t}.
\label{eq:schedule_increment}
\end{equation}
The CD-only evolution is therefore naturally discretized in the interpolation parameter \(s\). At fixed \(N_t\), changing \(T\) changes \(\dot{s}\) and \(\Delta t\) inversely, leaving the integrated CD angle unchanged.

\begin{figure*}
    \centering
    \includegraphics[width=\linewidth]{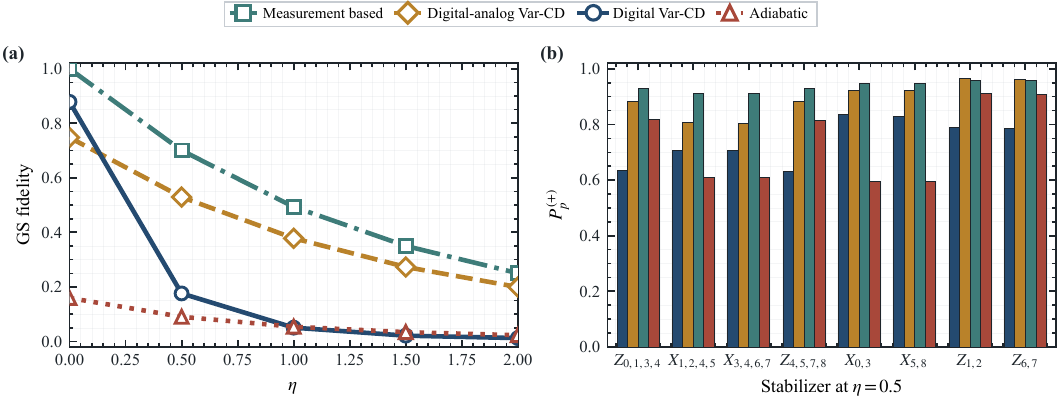}
    \caption{
    Robustness of logical-state preparation for the \(3\times3\) rotated surface-code patch.
    (a) Final ground-space fidelity \(F_{\mathrm{GS}}\) as a function of the dimensionless noise scale \(\eta\). The four curves correspond to Var-CD digital, Var-CD DAQC, single-shot stabilizer projection, and bare digital adiabatic evolution.
    (b) Plaquette-check probabilities \(P_p^{(+)}\) at \(\eta=0.5\).
    }
    \label{fig:noisy_hardware_simulation}
\end{figure*}

\subsection{Simulations with noise}
\label{sec:noisesimulation}

We investigate the robustness of the proposed preparation protocols using a hardware-motivated density-matrix noise model. The complete specification of the noise channels, native-operation durations, and reference error probabilities is given in Supplementary Sec.~\ref{sec:supp_noise_model}. We consider the \(3\times3\) rotated surface-code patch comparing digital Var-CD, digital-analog Var-CD, bare digital adiabatic evolution, and measurement-based preparation with syndrome-dependent Pauli-frame correction. All coherent protocols use the same total dimensionless evolution time, $T=1$,
and ten first-order Trotter steps, so $\Delta t=0.1$.

The strength of the hardware noise is controlled by a dimensionless parameter \(\eta\), which uniformly rescales the reference gate-error and decoherence parameters. The limit \(\eta=0\) corresponds to the ideal coherent evolution without noise.

For \(\eta>0\), the protocols are propagated as density matrices. Starting from
\begin{equation}
\rho(0)=\ket{\psi_0}\!\bra{\psi_0},
\end{equation}
the final ground-space fidelity is evaluated as
\begin{equation}
F_{\mathrm{GS}}
=
\mathrm{Tr}\!\left[
\rho(T)P_{\mathrm{GS}}
\right],
\label{eq:noisy_gs_fidelity}
\end{equation}
where \(P_{\mathrm{GS}}\) denotes the projector onto the degenerate ground space of the target stabilizer Hamiltonian. We additionally evaluate the probability that each stabilizer \(S_p\) is found in its desired \(+1\) eigenspace,
\begin{equation}
P_p^{(+)}
=
\frac{1+\langle S_p\rangle}{2},
\qquad
\langle S_p\rangle
=
\mathrm{Tr}\!\left[
\rho(T)S_p
\right].
\label{eq:noisy_stabilizer_probability}
\end{equation}
The local quantities \(P_p^{(+)}\) complement the global ground-space fidelity by resolving which stabilizer constraints are preserved or violated by the noisy preparation.

The noise parameters define a representative high-performance superconducting-qubit regime rather than a specific processor. At the reference noise scale \(\eta=1\), we take
\begin{equation}
p_{\mathrm{1q}}=2.0\times10^{-4},
\qquad
p_{\mathrm{2q}}=2.0\times10^{-3},
\end{equation}
corresponding to gate fidelities of \(99.98\%\) and \(99.8\%\), respectively. Single-qubit errors below \(10^{-4}\) have been demonstrated in transmons~\cite{Li2023SingleQubitError}, while superconducting CZ and iSWAP gates with fidelities near or above \(99.8\%\) have been reported using tunable couplers~\cite{Sung2021HighFidelity}. We assign the dressed analog blocks the same reference error scale,
\begin{equation}
p_{\mathrm{an}}=2.0\times10^{-3},
\end{equation}
as a modeling assumption corresponding to a calibrated analog block with an error comparable to a high-fidelity entangling operation.

We further use
\begin{equation}
T_1=150~\mu\mathrm{s},
\qquad
T_2^{*}=100~\mu\mathrm{s},
\end{equation}
which represent an optimistic but experimentally accessible high-coherence regime, superconducting devices with coherence times exceeding \(100~\mu\mathrm{s}\), and in some cases several hundred microseconds, have been demonstrated~\cite{Nguyen2019Fluxonium,Verjauw2022ManufacturableQubits}. The single- and two-qubit operation times are
\begin{equation}
t_{\mathrm{1q}}=25~\mathrm{ns},
\qquad
t_{\mathrm{2q}}=70~\mathrm{ns}.
\end{equation}
compatible with experimentally demonstrated superconducting gate times ranging from tens of nanoseconds to approximately \(100~\mathrm{ns}\)~\cite{Zhang2021FastFluxControl,Xu2020TunableCoupler}. The parameter \(\eta\) rescales the error probabilities, while the gate durations and coherence times remain fixed.

For the digital protocols, the implementation includes the required basis transformations, a parity-accumulation circuit, a single-qubit \(R_z\) rotation, and the inverse parity circuit. The resulting single- and two-qubit operation counts determine the effective depolarizing error and physical duration assigned to each Trotter step. Consequently, the digital Var-CD implementation incurs a larger noise cost than bare adiabatic evolution because each additional gauge-potential term must be synthesized as a separate many-qubit Pauli evolution.

The digital Var-CD Hamiltonian is evaluated independently at every Trotter midpoint,
\begin{equation}
H_{\mathrm{VarCD}}(\lambda_k)
=
H(\lambda_k)
+
\dot{\lambda}_k
A_{\lambda_k}^{\mathrm{Var}},
\end{equation}
where
\begin{equation}
A_{\lambda}^{\mathrm{Var}}
=
\sum_{\mu}
\alpha_{\mu}(\lambda)O_{\mu}.
\end{equation}
Its first-order digitized evolution is therefore
\begin{equation}
U_{\mathrm{VarCD}}(T)
\simeq
\prod_{k=1}^{N_t}
\prod_{\nu}
\exp\!\left[
-i\Delta t\,
h_{\nu}^{(k)}
\right],
\label{eq:noisy_digital_varcd_evolution}
\end{equation}
where \(h_{\nu}^{(k)}\) denotes an individual Pauli contribution to
\(H_{\mathrm{VarCD}}(\lambda_k)\).

The digital-analog protocol instead implements a fixed variational CD operator obtained at
\begin{equation}
\lambda_{\mathrm{fit}}=0.5.
\end{equation}
The corresponding coefficients are determined once and then used throughout the evolution. The interpolation Hamiltonian is not applied in the digital-analog branch, it is used only to construct the variational gauge potential and determine the fitted CD coefficients. The implemented digital-analog evolution is therefore
\begin{equation}
U_{\mathrm{DAQC}}(T)
\simeq
\prod_{k=1}^{N_t}
U_{\mathrm{CD}}^{(k)},
\label{eq:fixed_daqc_cd_evolution}
\end{equation}
with
\begin{equation}
U_{\mathrm{CD}}^{(k)}
\simeq
\exp\!\left[
-i\Delta t\,
\dot{\lambda}_k
A_{\lambda_{\mathrm{fit}}}^{\mathrm{DAQC}}
\right].
\end{equation}
For the linear schedule used here, \(\dot{\lambda}_k=1/T\) is constant.

The two-body boundary contributions to the CD operator are synthesized using dressed nearest-neighbor \(XY\) interactions. The higher-weight bulk contributions are generated using matching-based dressed analog blocks. For a selected block \(b\), the implemented unitary has the form
\begin{equation}
U_b
=
D_b^{\dagger}
\exp\!\left[
-i\Delta t
\sum_{(i,j)\in E_b}
J_{ij}^{(b)}
\left(
X_iX_j+Y_iY_j
\right)
\right]
D_b,
\label{eq:noisy_dressed_xy_block}
\end{equation}
where \(D_b\) denotes the local dressing transformation and the couplings \(J_{ij}^{(b)}\) are obtained by matching the desired CD coefficients. Only edges satisfying
\begin{equation}
\left|
\Delta t\,
\dot{\lambda}_k
J_{ij}^{(b)}
\right|
>
\epsilon_J
\end{equation}
are considered active analog interactions. Analog entangling noise is therefore applied only to the couplings that survive the synthesis threshold. Qubits participating in the dressing operations receive the corresponding control errors, while decoherence is included during the physical duration of the analog block.

The measurement-based reference consists of one round of stabilizer measurement followed by syndrome-dependent Pauli-frame correction. All syndrome outcomes are retained and corrected, consequently, the protocol is deterministic rather than postselected, and
\begin{equation}
P_{\mathrm{acc}}=1.
\end{equation}
In the ideal limit, each corrected syndrome branch is mapped into the simultaneous \(+1\) stabilizer sector, giving
\begin{equation}
F_{\mathrm{GS}}^{\mathrm{meas}}(\eta=0)=1.
\end{equation}
The implementation explicitly verifies the completeness and orthogonality of the syndrome branches, the corrected stabilizer eigenvalues, and the ground-space fidelity of every nonzero-probability corrected branch.

The measurement branch is evaluated by first constructing the exact syndrome-resolved projected states. These pure-state branches are subsequently converted into density matrices and combined according to their Born probabilities and the probabilities of syndrome-bit errors. The native resource cost of the stabilizer-extraction circuit is then included through an effective depolarizing channel and decoherence over the compiled circuit duration. The measurement-specific error model additionally contains ancilla-preparation, reset, and readout probabilities,
\begin{equation}
p_{\mathrm{prep}}=2\times10^{-3},
\quad
p_{\mathrm{reset}}=5\times10^{-3},
\quad
p_{\mathrm{readout}}=15\times10^{-3},
\end{equation}
at \(\eta=1\). The measurement and reset durations are both taken to be \(500~\mathrm{ns}\). Thus, the measurement curve represents an effective density-matrix model of stabilizer measurement and correction rather than a gate-by-gate noisy simulation of every ancilla operation.

Figure~\ref{fig:noisy_hardware_simulation}(a) shows the final ground-space fidelity as a function of \(\eta\). In the ideal limit, measurement and syndrome correction attains unit fidelity, as expected from exact stabilizer projection and correction. Among the coherent protocols, digital Var-CD gives the largest ideal fidelity. The digital-analog implementation has a lower ideal fidelity because it uses a fixed variational operator, a restricted hardware-native synthesis, and a CD-only evolution rather than the complete adiabatic-plus-CD Hamiltonian. Bare digital adiabatic evolution gives the lowest ideal fidelity because the total evolution time is too short to remain close to the instantaneous ground space.

The ordering changes as noise is increased. The fully digital Var-CD fidelity decreases rapidly because its ideal performance is obtained at the cost of a large number of decomposed Pauli-string evolutions. The associated basis transformations and entangling gates accumulate errors over the ten Trotter steps. Bare adiabatic evolution also decreases with increasing \(\eta\), although its performance is already limited by finite-time diabatic transitions in the noiseless limit.

The digital-analog Var-CD protocol is substantially more robust than the fully digital coherent protocols. Although its ideal fidelity is lower than that of digital Var-CD, it retains a considerably larger fraction of its ideal performance at finite noise. This behavior results from replacing long digital decompositions of the many-body CD terms by simultaneous dressed analog interactions. The analog implementation therefore trades some ideal synthesis accuracy for a lower effective operation count and improved robustness.

Measurement and syndrome correction gives the highest ground-space fidelity over the investigated noise range. Its advantage follows from the nonunitary stabilizer projection and syndrome-dependent correction, which actively transfer the state into the target stabilizer sector. This performance requires ancillas, mid-circuit measurements, reset operations, and classical Pauli-frame processing. In contrast, the digital-analog protocol is fully coherent and requires none of these measurement resources. The measurement-based protocol achieves the highest fidelity, but requires ancillas, mid-circuit measurements, reset operations, and classical Pauli-frame processing. The DAQC protocol instead provides a fully coherent and measurement-free preparation.

Figure~\ref{fig:noisy_hardware_simulation}(b) shows the stabilizer-resolved probabilities \(P_p^{(+)}\) at \(\eta=0.5\). Measurement and syndrome correction gives the largest and most uniform check probabilities, consistent with its explicit projection into the stabilizer sector. Digital-analog Var-CD also retains high probabilities for most bulk and boundary stabilizers, whereas the digital Var-CD probabilities are substantially reduced by the accumulated gate errors. Bare adiabatic evolution displays a pronounced dependence on stabilizer type. In particular, the initial computational-basis state already satisfies the \(Z\)-type constraints, while the \(X\)-type stabilizer correlations must be generated dynamically. This produces comparatively high \(Z\)-check probabilities and significantly weaker \(X\)-check probabilities at short evolution times.

The individual stabilizer probabilities should not be identified directly with the global ground-space fidelity. Even when every marginal probability \(P_p^{(+)}\) is relatively large, violations may occur in different stabilizer sectors across the mixed-state ensemble. The probability of satisfying all stabilizers simultaneously can therefore be considerably smaller than any individual \(P_p^{(+)}\). The combined analysis of \(F_{\mathrm{GS}}\) and \(P_p^{(+)}\) shows that the digital-analog protocol preserves both the global code-space overlap and the local stabilizer correlations more effectively than the fully digital coherent implementations at finite noise.

The present simulations compare noisy state-preparation protocols and do not model fault-tolerant quantum error correction. Only one round of stabilizer measurement and correction is included, and no repeated syndrome-extraction cycles or fault-tolerant decoder are simulated. Moreover, the measurement-circuit noise is incorporated through an effective resource-based channel rather than a complete gate-resolved ancilla simulation. Within these assumptions, measurement and syndrome correction provides the highest overall fidelity, while DAQC Var-CD is the most noise-robust fully coherent protocol. It therefore offers a measurement-free preparation strategy that avoids ancillas, mid-circuit measurements, reset operations, and real-time classical feedforward while retaining substantial ground-space fidelity under hardware-motivated noise.

\section{Discussion and Conclusion}
\label{sec:discussion_conclusion}

The central idea of this work is to reverse the conventional workflow for counterdiabatic state preparation. Rather than first constructing a variational adiabatic gauge potential (AGP) and subsequently compiling the resulting many-body Pauli operators into native gates, we first identify the many-body operator manifold that can be generated directly from the available hardware interactions. This hardware-generated manifold then serves as the variational ansatz for the AGP.

For the \(3\times3\) surface-code patch, local basis transformations of the native \(XY\) interaction generate the required two-body CD operators, while matching-based dressing generates the four-body operator families supported on the plaquette geometry. The union of these operators defines the variational AGP ansatz. Solving the Sels-Polkovnikov variational equations determines the corresponding CD coefficients, which are then used to set the native exchange couplings \(J_{ij}\) of the selected dressed analog blocks.

The synthesis depends on the local stabilizer geometry rather than on the total system size. Once the elementary dressing primitives are identified on the \(3\times3\) patch, the construction extends naturally to rectangular \(L_x\times L_y\) lattices by finding the matching patterns therefore avoiding many-body decomposition for each lattice.

The numerical benchmarks show that the resulting digital-analog protocol retains much of the advantage of digital Var-CD while employing a substantially shallower hardware-native entangling structure. The reduced digital gate overhead also improves robustness under the hardware-motivated noise model considered here. In particular, the digital-analog implementation maintains substantially higher ground-space fidelity than the fully digital coherent protocols and approaches the performance of the measurement-based preparation protocol over the investigated noise range. These results indicate that organizing the CD construction around the native hardware interaction can preserve the relevant stabilizer correlations while substantially reducing the cost of their implementation.

Although the present protocol addresses state preparation of the surface code rather than quantum error correction itself, preparing a state satisfying the stabilizer constraints is a fundamental prerequisite for logical-state initialization. The proposed construction therefore provides a coherent and measurement-free route to hardware-native preparation of stabilizer-code states, establishing a direct connection between counterdiabatic control and the initialization requirements of stabilizer-based quantum error-correcting architectures. This work opens a path toward digital-analog quantum error correction, where counterdiabatic control and dressed analog blocks can support stabilizer-state preparation and multi-qubit stabilizer operations, while digital operations handle ancilla control, readout, and recovery.

\section{Outlook}
\label{sec:outlook}

An important direction is to connect the present state-preparation protocol to a complete quantum-error-correction workflow. The method prepares states within the stabilizer ground-space manifold without relying on intermediate measurements, making it a natural candidate for the initialization stage preceding syndrome extraction. A next step is therefore to combine the counterdiabatic preparation with repeated stabilizer measurements, decoding, and recovery operations, and to quantify how the initial preparation fidelity propagates to logical observables and logical error rates. The same construction could also be explored for preparing specific logical states and for initialization of larger surface-code, toric-code, and other stabilizer-code architectures.

More broadly, the construction developed here is not restricted to stabilizer Hamiltonians. Its essential ingredient is the generation of a higher-order counterdiabatic operator manifold from native two-body interactions before performing the variational AGP optimization. In the present implementation, basis transformations and matching-based dressing generate operators up to four-body locality from the native \(XY\) interaction. Extending the dressing structure, matching patterns, or sequence depth could provide access to larger operator manifolds and higher-body interactions. This opens a route to applying the same strategy to other many-body problems in which the relevant counterdiabatic operators contain non-native multi-qubit terms, including lattice-gauge theories, higher-order spin models, and optimization problems with multi-body cost functions.

\begin{acknowledgments}
This work has been partially supported by the Valencian Government Grant No. CIAICO/2024/111 and the Spanish Ministry of Economic Affairs and Digital Transformation through the QUANTUM ENIA project call – Quantum Spain project, and the European Union through the Recovery, Transformation and Resilience Plan – NextGenerationEU (Digital Spain 2026) Agenda.
\end{acknowledgments}

\section*{Data Availability}

Data supporting the findings of this study are available from the corresponding author upon reasonable request.

\clearpage
\section*{Supplementary Material}
\addcontentsline{toc}{section}{Supplementary Material}

\begin{table*}
\centering
\renewcommand{\arraystretch}{1.15}
\setlength{\tabcolsep}{3.8pt}
\scriptsize
\caption{
CD coefficients and resource comparison for the \(3\times3\) checkerboard plaquette-stabilizer benchmark per Trotter step. Resources are measured in terms of the number of native two-qubit exchange gates (2QGs), single-qubit rotations (1QGs), and analog blocks (ABs). For the hybrid digital-analog implementation, six out of 12 four-body operators selected by the VarAGP are synthesized using dressed analog blocks, while the two-body boundary operators are implemented digitally and included in the reported resource counts.
}
\label{tab:3x3_coeff_structure}
\begin{tabular}{|l|l|c|c|c|c|}
\hline
\textbf{Method}
& \textbf{Pauli term}
& \textbf{Coefficient at \(s=0.5\)}
& \textbf{Depth / step}
& \textbf{2QGs or ABs / step}
& \textbf{1QGs / step} \\
\hline

Digital-analog Var-CD
& \(Y_{1}X_{2}X_{4}X_{5}\) & \(0.2308\) & \(25\)
& \(8\) native \(XY\) gates \(+\ 4\) ABs & \(28\) explicit \\
& \(X_{1}Y_{2}X_{4}X_{5}\) & \(0.2308\) & & & \\
& \(X_{1}X_{2}Y_{4}X_{5}\) & \(0.1154\) & & & \\
& \(X_{1}X_{2}X_{4}Y_{5}\) & \(0.0769\) & & & \\
& \(Y_{3}X_{4}X_{6}X_{7}\) & \(0.0769\) & & & \\
& \(X_{3}Y_{4}X_{6}X_{7}\) & \(0.1154\) & & & \\
& \(Y_{0}X_{3}\) & \(0.5385\) & & & \\
& \(X_{0}Y_{3}\) & \(0.1923\) & & & \\
& \(Y_{5}X_{8}\) & \(0.1923\) & & & \\
& \(X_{5}Y_{8}\) & \(0.5385\) & & & \\
\hline

Digital Var-CD
& \(Y_{1}X_{2}X_{4}X_{5}\) & \(0.2308\) & \(211\)
& \(92\) iswap gates & \(305\) \\
& \(X_{1}Y_{2}X_{4}X_{5}\) & \(0.2308\) & & & \\
& \(X_{1}X_{2}Y_{4}X_{5}\) & \(0.1154\) & & & \\
& \(X_{1}X_{2}X_{4}Y_{5}\) & \(0.0769\) & & & \\
& \(Y_{3}X_{4}X_{6}X_{7}\) & \(0.0769\) & & & \\
& \(X_{3}Y_{4}X_{6}X_{7}\) & \(0.1154\) & & & \\
& \(X_{3}X_{4}Y_{6}X_{7}\) & \(0.2308\) & & & \\
& \(X_{3}X_{4}X_{6}Y_{7}\) & \(0.2308\) & & & \\
& \(Y_{0}X_{3}\) & \(0.5385\) & & & \\
& \(X_{0}Y_{3}\) & \(0.1923\) & & & \\
& \(Y_{5}X_{8}\) & \(0.1923\) & & & \\
& \(X_{5}Y_{8}\) & \(0.5385\) & & & \\
\hline

Digital NC-CD
& \(Y_{1}X_{2}X_{4}X_{5}\) & \(0.1905\) & \(238\)
& \(108\) iswap gates & \(340\) \\
& \(X_{1}Y_{2}X_{4}X_{5}\) & \(0.1905\) & & & \\
& \(X_{1}X_{2}Y_{4}X_{5}\) & \(0.1905\) & & & \\
& \(X_{1}X_{2}X_{4}Y_{5}\) & \(0.1905\) & & & \\
& \(Y_{3}X_{4}X_{6}X_{7}\) & \(0.1905\) & & & \\
& \(X_{3}Y_{4}X_{6}X_{7}\) & \(0.1905\) & & & \\
& \(X_{3}X_{4}Y_{6}X_{7}\) & \(0.1905\) & & & \\
& \(X_{3}X_{4}X_{6}Y_{7}\) & \(0.1905\) & & & \\
& \(Y_{0}X_{3}\) & \(0.1905\) & & & \\
& \(X_{0}Y_{3}\) & \(0.1905\) & & & \\
& \(Y_{5}X_{8}\) & \(0.1905\) & & & \\
& \(X_{5}Y_{8}\) & \(0.1905\) & & & \\
\hline
\end{tabular}
\end{table*}

\begin{table*}
\centering
\renewcommand{\arraystretch}{1.12}
\setlength{\tabcolsep}{5pt}
\caption{Dressed two-edge matchings used for the \(3\times3\) DAQC-CD construction.}
\label{tab:3x3_dressing_matchings}
\begin{tabular}{cll}
\hline
Matching label & Matched edges & Two-qubit dressings ($D_{mv}$)\\
\hline
M1
& \(\{(1,2),(4,5)\}\)
& \((1,2):\,XX+ZY,\qquad (4,5):\,ZX+XY\)
\\
M2
& \(\{(1,2),(4,5)\}\)
& \((1,2):\,XX+YZ,\qquad (4,5):\,XZ+YX\)
\\
M3
& \(\{(3,4),(6,7)\}\)
& \((3,4):\,XX+ZY,\qquad (6,7):\,ZX+XY\)
\\
M4
& \(\{(3,4),(6,7)\}\)
& \((3,4):\,XX+YZ,\qquad (6,7):\,XZ+YX\)
\\
\hline
\end{tabular}
\end{table*}

\begin{table}[t]
\centering
\renewcommand{\arraystretch}{1.15}
\begin{tabular}{ccl}
\hline\hline
Edge & Generated Pauli & Type\\
\hline
\((4,5)\) &
\(Y_1X_2X_4X_5+X_1Y_2X_4X_5\) &
target CD terms\\
\((3,4)\) &
\(Y_0X_1X_3X_4+X_0Y_1X_3X_4\) &
target CD terms\\
\((0,1)\) &
\(-X_3X_4+Z_0Z_1Z_3Z_4\) &
residual/stabilizer terms\\
\((1,2)\) &
\(-X_4X_5+Z_1Z_2Z_4Y_5\) &
residual terms\\
\hline\hline
\end{tabular}
\caption{Representative Pauli terms generated by dressing individual nearest-neighbor XY edges in the \(3\times3\) example. The edges are shown separately only to identify their contributions, all nonzero \(J_{ij}\) couplings are executed simultaneously within an analog block.}
\label{tab:dressed_edge_examples}
\end{table}

\section{Digital-analog compilation of the variational counterdiabatic protocol}
\label{app:daqc_compilation}

The digital-analog implementation exploits the native nearest-neighbor exchange interaction while preserving the variational counterdiabatic (Var-CD) operator obtained from the plaquette-local ansatz. Rather than digitally decomposing every Pauli string, the four-body CD operators are synthesized using basis-transformed analog exchange blocks, whereas the remaining two-body boundary terms are implemented digitally.

\subsection{Higher-order decomposition}
\label{subsec:higher_order_interactions}

The digital-analog construction starts from the native nearest-neighbor exchange Hamiltonian
\begin{equation}
H_{\rm XY}(\mathbf J)=\sum_{(i,j)\in E}J_{ij}(X_iX_j+Y_iY_j).
\label{eq:xy_native}
\end{equation}
For a matching \(m\), the dressing circuit is
\begin{equation}
D_{m,\nu}
=
R_\nu
\prod_{(a,b)\in m}
e^{-i\frac{\pi}{4}(X_aX_b+Y_aY_b)},
\label{eq:dressing}
\end{equation}
where the product represents fixed-angle iSWAP-type interactions applied in parallel on the disjoint edges of \(m\), while \(R_\nu\) contains the single-qubit rotations defining the basis choice \(\nu\). For the basis choices used here, local \(\pi/2\) rotations ($R_\nu$) about the \(x\)- and \(y\)-axes transform the native \(XY\) interaction into the dressed two-qubit generators
\begin{equation}
Z_aX_b+X_aY_b,
\qquad
X_aX_b+Z_aY_b,
\label{eq:dressing_gates}
\end{equation}
together with their qubit-swapped variants. The dressed analog evolution is
\begin{equation}
U_{\rm DAQC}
=
D_{m,\nu}^{\dagger}
e^{-itH_{\rm XY}(\mathbf J)}
D_{m,\nu},
\label{eq:daqc_basis}
\end{equation}
Different choices of the matching \(m\) and basis \(\nu\) therefore generate different dressed Pauli operators from the same native \(XY\) interaction. Because the two-qubit dressing operations act on disjoint edges of the matching, they can be applied in parallel. Conjugating the remaining exchange couplings by this dressing generates higher-weight Pauli strings. Different choices of \(m\) and \(\nu\) therefore generate different two-, three-, and four-body operators from the same native \(XY\) interaction. For example, an appropriate matching and local Pauli frame can transform a native exchange edge according to
\begin{equation}
D^\dagger
\left(X_4X_5+Y_4Y_5\right)
D
=
Y_1X_2X_4X_5
+
X_1Y_2X_4X_5,
\label{eq:example_four_body_generation}
\end{equation}
thereby producing two of the four-body single-\(Y\) operators required by the Var-CD ansatz.

For the \(3\times3\) geometry shown in Fig.~\ref{fig:daqc_model}, the full dressed interaction contains a combination of one-, two-, three-, and four-body operators. By switching off the couplings that do not contribute to the desired plaquette structure, the interaction reduces to
\begin{equation}
\begin{aligned}
H_{\rm eff}={}&
J_{5,8}
\left(
X_2Z_5Z_7Z_8
+
Z_2Z_5X_7Z_8
\right)\\
&+
J_{4,5}
\left(
X_1Z_2Z_4Z_5
+
Z_1X_2Z_4Z_5
\right)\\
&+
J_{3,4}
\left(
X_0Z_1Z_3Z_4
+
Z_0X_1Z_3Z_4
\right).
\end{aligned}
\label{eq:selected_four_body_terms}
\end{equation}
A further local \(R_z(\pi/2)\) basis rotation maps \(X\rightarrow Y\), giving
\begin{equation}
\begin{aligned}
H_{\rm eff}={}&
J_{5,8}
\left(
Y_2Z_5Z_7Z_8
+
Z_2Z_5Y_7Z_8
\right)\\
&+
J_{4,5}
\left(
Y_1Z_2Z_4Z_5
+
Z_1Y_2Z_4Z_5
\right)\\
&+
J_{3,4}
\left(
Y_0Z_1Z_3Z_4
+
Z_0Y_1Z_3Z_4
\right).
\end{aligned}
\label{eq:selected_four_body_terms_y}
\end{equation}

Thus, the same programmable \(XY\) interaction can generate different many-body Pauli structures through the choice of matching, local basis rotations, and exchange couplings. The fitted \(J_{ij}\) determine the amplitudes of the generated operators, while the dressing circuit determines their operator structure. This provides the hardware-native operator pool used to construct the variational counterdiabatic protocol.

The growth of the operator weight under the dressing can be understood directly from its support. Let \(\mathrm{wt}(P)\) denote the Pauli weight of \(P\), i.e., the number of qubits on which \(P\) acts nontrivially. Since each dressing layer consists of two-qubit gates acting on disjoint pairs, every qubit in the support of \(P\) can introduce at most one additional qubit under conjugation. Therefore,
\begin{equation}
\mathrm{wt}\!\left(D^\dagger P D\right)
\leq
\min\!\left(N,2\,\mathrm{wt}(P)\right).
\label{eq:weight_growth_single}
\end{equation}
Since each native exchange term has initial weight \(\mathrm{wt}(P_0)=2\), a single dressing layer can generate operators with weight up to four, as explicitly demonstrated above. More generally, after \(\ell\) successive dressing layers,
\begin{equation}
\mathrm{wt}(P_\ell)
\leq
\min\!\left(N,2^\ell\mathrm{wt}(P_0)\right)
=
\min\!\left(N,2^{\ell+1}\right).
\label{eq:weight_growth}
\end{equation}
Consequently, the four-body interactions required here are generated using a single layer of parallel two-qubit dressing gates, followed by the analog evolution and the inverse dressing. For the fixed four-body operator weight considered in this work, the resulting digital-analog construction therefore has constant circuit depth with increasing system size \(N\). The caveat is that this doesn´t generate all the terms for a given Hamiltonian, and therefore demands enumerating over all relevant matching pairs. 

An important feature of this construction is that the generated higher-body interactions remain controlled through the underlying pairwise exchange couplings \(J_{ij}\), rather than requiring directly tunable four-body couplings \(J_{ijkl}\). The amplitudes of the dressed Pauli operators are therefore determined by experimentally accessible two-qubit coupling parameters. In contrast to approaches that rely on directly engineered many-body interactions \cite{Richerme.2025,menke2022threebody,busnaina2025lattice}, the present construction uses pairwise interactions together with digital dressing to synthesize structured higher-body Hamiltonians. This provides a flexible route to engineering the four-body interactions required for the surface-code counterdiabatic protocol while retaining pairwise coupling control.
\subsection{Parallel matching construction}

The dressing transformations are applied only on disjoint nearest-neighbor pairs. A matching
\begin{equation}
m=\{(i_1,j_1),\ldots,(i_r,j_r)\}
\end{equation}
contains no qubit more than once, allowing all corresponding two-qubit basis transformations to be executed simultaneously.

The matching pool is generated using a brute-force search over the nearest-neighbor edges of the lattice. All edge subsets containing up to four edges are enumerated, and subsets in which two edges share a qubit are discarded. The remaining subsets define the valid matchings.

For the \(3\times3\) lattice, all valid matchings with up to four disjoint edges are generated in this way. Each matching is combined with four local basis choices to construct candidate dressed analog blocks. Conjugating the native exchange Hamiltonian by each candidate gives
\begin{equation}
\mathcal P=\{D^\dagger H_{\rm XY}D\},
\end{equation}
from which the operators belonging to the Var-CD basis are identified.

\subsection{Transfer of matching patterns to the \(4\times4\) lattice}
\label{sec:matching_transfer}

For the \(4\times4\) lattice, we reuse the matching structures selected on the \(3\times4\) reference system. The \(3\times4\) search contains 302 maximum-size matchings, but the final DAQC construction uses only two, which reduce to one local matching template after removing their absolute lattice positions. We generate all valid translated, rotated, and reflected copies of this template on the \(4\times4\) lattice.

The resulting transferred pool contains only six matchings and still covers all 16 required four-body CD terms. Table~\ref{tab:matching_transfer} compares this construction with a complete \(4\times4\) matching search.

\begin{table}[t]
\centering
\caption{Comparison of the full and transferred matching pools for the \(4\times4\) lattice.}
\begin{tabular}{lcc}
\hline
Quantity & Brute force & Transferred pool\\
\hline
Candidate matchings & 2593 & 6\\
Surviving dressed blocks & 1096 & 12\\
Required four-body CD terms & 16 & 16\\
Covered four-body CD terms & 16 & 16\\
Missing four-body CD terms & 0 & 0\\
\hline
\end{tabular}
\label{tab:matching_transfer}
\end{table}

Thus, the transfer reduces the matching pool from 2593 to 6 candidates while preserving the complete four-body CD-operator coverage. Only the matching structure is transferred, the VarCD coefficients and the corresponding couplings \(J_{ij}\) are recomputed for the \(4\times4\) Hamiltonian.

\subsection{Digital-analog counterdiabatic evolution}

The fully digital Var-CD protocol implements the counterdiabatic Hamiltonian
\begin{equation}
H_{\rm CD}(\lambda)
=
H(\lambda)
+
\dot{\lambda}A(\lambda),
\label{eq:cd_hamiltonian_appendix}
\end{equation}
where the gauge potential is expanded as
\begin{equation}
A(\lambda)
=
\sum_{\mu}\alpha_\mu(\lambda)O_\mu,
\label{eq:var_basis}
\end{equation}
with the coefficients obtained from the Sels-Polkovnikov variational principle. For the \(3\times3\) rotated surface-code patch, the operator basis contains twelve terms, consisting of four two-body boundary operators and eight four-body plaquette operators.

For the digital-analog implementation, the Var-CD coefficients are evaluated once at the beginning and kept fixed throughout the subsequent time evolution.

The corresponding digital-analog CD evolution is implemented as a product of the selected dressed analog blocks,
\begin{equation}
U_{\rm DAQC}
=
\prod_{(m,\nu)}
D_{m,\nu}^{\dagger}
e^{-itH_{\rm XY}(\mathbf J)}
D_{m,\nu}.
\label{eq:daqc_product}
\end{equation}
Here, \(m\) specifies the matching and \(\nu\) the corresponding basis choice, while the exchange couplings \(J_{ij}\) and evolution times \(t_{m,\nu}\) determine the amplitudes of the generated Pauli operators.

\section{Noise model}
\label{sec:supp_noise_model}

The noisy simulations are performed at the density-matrix level using a hardware-motivated effective noise model. A dimensionless parameter \(\eta\) uniformly rescales the baseline error probabilities, with \(\eta=0\) corresponding to ideal evolution and \(\eta=1\) defining the reference hardware setting. Values \(\eta>1\) probe stronger noise levels. The stabilizer probabilities shown in Fig.~\ref{fig:noisy_hardware_simulation}(b) are evaluated at \(\eta=0.5\).

For the coherent preparation protocols, the ideal unitary associated with each operation is followed by depolarizing noise, amplitude damping, and pure dephasing. For the digital adiabatic and digital Var-CD protocols, the effective noise strength is determined from the resource profile of the Pauli-string decomposition. The accumulated depolarizing probability assigned to qubit \(q\) is
\begin{equation}
p_q(\eta)
=
1-
(1-\eta p_{1q})^{n_{1q}^{(q)}}
(1-\eta p_{2q})^{n_{2q}^{(q)}},
\label{eq:supp_digital_depolarizing_probability}
\end{equation}
where \(n_{1q}^{(q)}\) and \(n_{2q}^{(q)}\) denote the effective numbers of single- and two-qubit operations acting on qubit \(q\), respectively. The physical duration assigned to each digital evolution is obtained from the corresponding single- and two-qubit gate counts and is bounded from below by the reference physical duration of one Trotter interval.

The digital-analog Var-CD protocol implements only the counterdiabatic contribution using locally dressed exchange blocks. Depolarizing errors are assigned to the single- and two-qubit operations forming the dressing transformations. Analog-block errors are applied only to synthesized couplings whose integrated interaction angles satisfy the activity threshold
\begin{equation}
\left|\theta_{ij}\right|
=
\left|
\Delta t\,\dot{\lambda}\,J_{ij}
\right|
>
\epsilon_J .
\end{equation}
The total analog-block error is distributed among the active couplings according to their relative integrated interaction angles. Thus, inactive edges are not assigned an artificial entangling error. The physical duration of an analog block is evaluated as
\begin{equation}
t_{\mathrm{an}}
=
\min\!\left[
\frac{\max_{ij}|\theta_{ij}|}{g},
t_{\mathrm{an}}^{\max}
\right],
\label{eq:supp_analog_duration}
\end{equation}
where \(g\) is the reference exchange-coupling strength.

The single-qubit depolarizing channel is
\begin{equation}
\mathcal{E}_{1q}(\rho)
=
(1-p)\rho
+
\frac{p}{3}
\sum_{\sigma\in\{X,Y,Z\}}
\sigma\rho\sigma,
\label{eq:supp_single_qubit_depolarizing}
\end{equation}
while the two-qubit depolarizing channel is
\begin{equation}
\mathcal{E}_{2q}(\rho)
=
(1-p)\rho
+
\frac{p}{15}
\sum_{\substack{P_i,P_j\in\{I,X,Y,Z\}\\
(P_i,P_j)\neq(I,I)}}
(P_i\otimes P_j)\rho(P_i\otimes P_j).
\label{eq:supp_two_qubit_depolarizing}
\end{equation}

Amplitude damping and pure dephasing are applied during every operation of physical duration \(t\). Their probabilities are
\begin{equation}
p_{\mathrm{amp}}(t,\eta)
=
\eta\left(1-e^{-t/T_1}\right),
p_{\phi}(t,\eta)
=
\frac{\eta}{2}
\left(1-e^{-t/T_{\phi}}\right),
\label{eq:supp_decoherence_probabilities}
\end{equation}
with
\begin{equation}
\frac{1}{T_{\phi}}
=
\frac{1}{T_2^{*}}
-
\frac{1}{2T_1}.
\label{eq:supp_pure_dephasing_time}
\end{equation}
Amplitude damping is implemented through the Kraus operators
\begin{equation}
K_0
=
\begin{pmatrix}
1 & 0\\
0 & \sqrt{1-p_{\mathrm{amp}}}
\end{pmatrix},
\qquad
K_1
=
\begin{pmatrix}
0 & \sqrt{p_{\mathrm{amp}}}\\
0 & 0
\end{pmatrix},
\end{equation}
whereas pure dephasing is represented by the channel
\begin{equation}
\mathcal{E}_{\phi}(\rho)
=
(1-p_{\phi})\rho
+
p_{\phi}Z\rho Z.
\end{equation}

The measurement-based reference consists of a single round of ancilla-assisted stabilizer measurement followed by syndrome-dependent virtual Pauli-frame correction. All syndrome outcomes are retained and corrected, so the protocol is deterministic,
\begin{equation}
P_{\mathrm{acc}}=1,
\end{equation}
rather than postselected. Repeated syndrome extraction, decoding, and fault-tolerant correction are not included.

The exact syndrome-resolved branches are first constructed from the projected statevectors. These branches are converted into density matrices and combined according to their Born probabilities and the probabilities of syndrome-bit errors. Ancilla-preparation, readout, and reset errors are represented through the effective syndrome-bit error probability
\begin{equation}
p_{\mathrm{syn}}(\eta)
=
1-
(1-\eta p_{\mathrm{ro}})
(1-\eta p_{\mathrm{reset}})
(1-\eta p_{\mathrm{anc}}).
\label{eq:supp_syndrome_error}
\end{equation}
The resource cost of the compiled stabilizer-extraction circuit is subsequently incorporated through an effective depolarizing channel and decoherence over its total physical duration. The measurement-based model is therefore an effective resource-level description rather than a gate-resolved density-matrix simulation of every ancilla operation.

The reference parameters used in the simulations are summarized in Table~\ref{tab:supp_noise_parameters}. They define a representative high-performance superconducting-qubit regime rather than a specific processor. The dimensionless parameter \(\eta\) rescales the listed error probabilities, while the gate durations, coherence times, and coupling strength remain fixed.

\begin{table}[t]
\centering
\caption{Reference parameters of the effective hardware-motivated noise model used in the density-matrix simulations.}
\label{tab:supp_noise_parameters}
\begin{tabular}{lc}
\hline
Quantity & Value\\
\hline
Single-qubit gate error \(p_{1q}\)
& \(2.0\times10^{-4}\)\\
Two-qubit gate error \(p_{2q}\)
& \(2.0\times10^{-3}\)\\
Analog-block error \(p_{\mathrm{an}}\)
& \(2.0\times10^{-3}\)\\
Readout error \(p_{\mathrm{ro}}\)
& \(1.5\times10^{-2}\)\\
Ancilla-preparation error \(p_{\mathrm{anc}}\)
& \(2.0\times10^{-3}\)\\
Reset error \(p_{\mathrm{reset}}\)
& \(5.0\times10^{-3}\)\\
Relaxation time \(T_1\)
& \(150\,\mu\mathrm{s}\)\\
Ramsey dephasing time \(T_2^{*}\)
& \(100\,\mu\mathrm{s}\)\\
Single-qubit gate duration \(t_{1q}\)
& \(25\,\mathrm{ns}\)\\
Two-qubit/\(\mathrm{iSWAP}\) duration \(t_{2q}\)
& \(70\,\mathrm{ns}\)\\
Measurement duration \(t_{\mathrm{meas}}\)
& \(500\,\mathrm{ns}\)\\
Reset duration \(t_{\mathrm{reset}}\)
& \(500\,\mathrm{ns}\)\\
Maximum analog-block duration \(t_{\mathrm{an}}^{\max}\)
& \(100\,\mathrm{ns}\)\\
Reference coupling \(g/2\pi\)
& \(10\,\mathrm{MHz}\)\\
Reference physical evolution time \(t_{\mathrm{phys}}\)
& \(500\,\mathrm{ns}\)\\
\hline
\end{tabular}
\end{table}

\clearpage

\bibliography{ref}

@article{Kitaev2003,
  author  = {Kitaev, A. Yu.},
  title   = {Fault-tolerant quantum computation by anyons},
  journal = {Annals of Physics},
  volume  = {303},
  number  = {1},
  pages   = {2--30},
  year    = {2003},
  doi     = {10.1016/S0003-4916(02)00018-0},
  url = {https://doi.org/10.1016/S0003-4916(02)00018-0}
}

@article{Dennis2002,
  author  = {Dennis, Eric and Kitaev, Alexei and Landahl, Andrew and Preskill, John},
  title   = {Topological quantum memory},
  journal = {Journal of Mathematical Physics},
  volume  = {43},
  number  = {9},
  pages   = {4452--4505},
  year    = {2002},
  doi     = {10.1063/1.1499754},
  url = {https://doi.org/10.1063/1.1499754}
}

@misc{Farhi2000,
  author        = {Farhi, Edward and Goldstone, Jeffrey and Gutmann, Sam and Sipser, Michael},
  title         = {Quantum Computation by Adiabatic Evolution},
  year          = {2000},
  eprint        = {quant-ph/0001106},
  archivePrefix = {arXiv},
  primaryClass  = {quant-ph},
  doi           = {10.48550/arXiv.quant-ph/0001106},
  url           = {https://arxiv.org/abs/quant-ph/0001106}
}

@article{Aharonov2008,
  author  = {Aharonov, Dorit and van Dam, Wim and Kempe, Julia and Landau, Zeph and Lloyd, Seth and Regev, Oded},
  title   = {Adiabatic Quantum Computation Is Equivalent to Standard Quantum Computation},
  journal = {SIAM Review},
  volume  = {50},
  number  = {4},
  pages   = {755--787},
  year    = {2008},
  doi     = {10.1137/080734479},
  url = {https://doi.org/10.1137/080734479}
}

@article{Fowler2012,
  author  = {Fowler, Austin G. and Mariantoni, Matteo and Martinis, John M. and Cleland, Andrew N.},
  title   = {Surface codes: Towards practical large-scale quantum computation},
  journal = {Physical Review A},
  volume  = {86},
  number  = {3},
  pages   = {032324},
  year    = {2012},
  doi     = {10.1103/PhysRevA.86.032324},
  url = {https://doi.org/10.1103/PhysRevA.86.032324}
}

@article{Richerme.2025,
  author = {Richerme, Philip},
  title = {Multi-Mode Global Driving of Trapped Ions for Quantum Circuit Synthesis},
  journal = {Quantum Science and Technology},
  volume = {10},
  pages = {035046},
  year = {2025},
  doi = {10.1088/2058-9565/ade560},
  url = {https://doi.org/10.1088/2058-9565/ade560}
}

@article{menke2022threebody,
  title={Demonstration of Tunable Three-Body Interactions Between Superconducting Qubits},
  author={Menke, Tim and Banner, William P. and Bergamaschi, Thomas R. and others},
  journal={Physical Review Letters},
  volume={129},
  number={22},
  pages={220501},
  year={2022},
  doi={10.1103/PhysRevLett.129.220501},
  url = {https://doi.org/10.1103/PhysRevLett.129.220501}
}

@article{busnaina2025lattice,
  author = {Busnaina, J. H. and Shi, Z. and Alcaine-Cuervo, Jes{\'u}s M. and Yang, Cindy X. and Nsanzineza, I. and Rico, E. and Wilson, C. M.},
  title = {Native Three-Body Interactions in a Superconducting Lattice Gauge Quantum Simulator},
  journal = {Physical Review B},
  volume = {112},
  pages = {134514},
  year = {2025},
  doi = {10.1103/6prx-zmdz},
  url = {https://doi.org/10.1103/6prx-zmdz}
}

@article{Lamata2018,
  author  = {Lucas Lamata and Adrian Parra-Rodriguez and Mikel Sanz and Enrique Solano},
  title   = {Digital‐Analog Quantum Simulations with Superconducting Circuits},
  journal = {Advances in Physics: X},
  volume  = {3},
  number  = {1},
  pages   = {1457981},
  year    = {2018},
  doi     = {10.1080/23746149.2018.1457981},
  url     = {https://doi.org/10.1080/23746149.2018.1457981}
}

@article{Mikel.2020,
  author = {A. Parra-Rodriguez and P. Lougovski and L. Lamata and E. Solano and M. Sanz},
  title = {Digital-analog quantum computations},
  journal = {Physical Review A},
  volume = {101},
  pages = {022305},
  year = {2020},
  url = {https://link.aps.org/doi/10.1103/PhysRevA.101.022305}
}

@article{Julen.2016,
title  = {Digital-Analog Quantum  Simulation of Spin Models in Trapped Ions},
Author = {Arrazola, Iñigo and Pedernales, Julen S. and Lamata, Lucas and Solano, Enrique},
Journal =  {Scientific Reports},
Year = {2016},
Volume = {6},
Pages = {30534},
url = {https://doi.org/10.1038/srep30534}
}

@article{Kumar_2025,
doi = {10.1088/2058-9565/ad8b64},
url = {https://dx.doi.org/10.1088/2058-9565/ad8b64},
year = {2024},
month = {nov},
publisher = {IOP Publishing},
volume = {10},
number = {1},
pages = {015023},
author = {Kumar, Shubham and Hegade, Narendra N and Henrique de Oliveira, Murilo and Solano, Enrique and Gomez Cadavid, Alejandro and Albarrán-Arriagada, F},
title = {Digital-analog counterdiabatic quantum optimization with trapped ions},
journal = {Quantum Science and Technology}
}

@Article{Kumar_fermion_boson2025,
author={Kumar, Shubham
and Hegade, Narendra N.
and Visuri, Anne-Maria
and Bhargava, Balaganchi A.
and Hernandez, Juan F. R.
and Solano, E.
and Albarr{\'a}n-Arriagada, F.
and Barrios, G. Alvarado},
title={Digital-analog quantum computing of fermion-boson models in superconducting circuits},
journal={npj Quantum Information},
year={2025},
month={Mar},
day={11},
volume={11},
number={1},
pages={43},
issn={2056-6387},
doi={10.1038/s41534-025-01001-4},
url={https://doi.org/10.1038/s41534-025-01001-4}
}

@misc{Bhargava2026,
      title={Constant Depth Digital-Analog Counterdiabatic Quantum Computing}, 
      author={Balaganchi A. Bhargava and Shubham Kumar and Anne-Maria Visuri and Paolo A. Erdman and Enrique Solano and Narendra N. Hegade},
      year={2026},
      eprint={2601.01154},
      archivePrefix={arXiv},
      primaryClass={quant-ph},
      url={https://arxiv.org/abs/2601.01154}, 
}

@article{Gonzalez-Raya2021DigitalAnalog,
  title        = {Digital-Analog Quantum Simulations Using the Cross-Resonance Effect},
  author       = {Gonzalez-Raya, Tasio and Asensio-Perea, Rodrigo and Martin, Ana and C{\'e}leri, Lucas~C. and Sanz, Mikel and Lougovski, Pavel and Dumitrescu, Eugene~F.},
  journal      = {PRX Quantum},
  volume       = {2},
  number       = {2},
  pages        = {020328},
  year         = {2021},
  publisher    = {American Physical Society},
  doi          = {10.1103/PRXQuantum.2.020328},
  url = {https://doi.org/10.1103/PRXQuantum.2.020328}
}

@article{Pedram.2024,
  author = {Andersen, T. I. and others},
  title = {Thermalization and Criticality on an Analogue-Digital Quantum Simulator},
  journal = {Nature},
  volume = {638},
  pages = {79--85},
  year = {2025},
  doi = {10.1038/s41586-024-08460-3},
  url = {https://doi.org/10.1038/s41586-024-08460-3}
}

@article{Arute2019,
  author = {Arute, Frank and others},
  title = {Quantum supremacy using a programmable superconducting processor},
  journal = {Nature},
  volume = {574},
  pages = {505--510},
  year = {2019},
  doi = {10.1038/s41586-019-1666-5},
  url = {https://doi.org/10.1038/s41586-019-1666-5}
}

@article{Gong.2023,
  author = {Gong, Ming and others},
  title = {Quantum neuronal sensing of quantum many-body states on a 61-qubit programmable superconducting processor},
  journal = {Science Bulletin},
  volume = {68},
  number = {9},
  pages = {906--912},
  year = {2023},
  doi = {10.1016/j.scib.2023.04.003},
  url = {https://doi.org/10.1016/j.scib.2023.04.003}
}

@article{PhysRevLett.123.090602,
  title = {Floquet-Engineering Counterdiabatic Protocols in Quantum Many-Body Systems},
  author = {Claeys, Pieter W. and Pandey, Mohit and Sels, Dries and Polkovnikov, Anatoli},
  journal = {Physical Review Letters},
  volume = {123},
  issue = {9},
  pages = {090602},
  numpages = {7},
  year = {2019},
  month = {Aug},
  publisher = {American Physical Society},
  doi = {10.1103/PhysRevLett.123.090602},
  url = {https://link.aps.org/doi/10.1103/PhysRevLett.123.090602}
}

@article{Sels2017,
  title   = {Minimizing irreversible losses in quantum systems by local counterdiabatic driving},
  author  = {Sels, Dries and Polkovnikov, Anatoli},
  journal = {Proceedings of the National Academy of Sciences},
  volume  = {114},
  number  = {20},
  pages   = {E3909--E3916},
  year    = {2017},
  doi     = {10.1073/pnas.1619826114},
  url     = {https://doi.org/10.1073/pnas.1619826114}
}

@article{PhysRevB.106.155153,
  title = {Variational counterdiabatic driving of the Hubbard model for ground-state preparation},
  author = {Xie, Qing and Seki, Kazuhiro and Yunoki, Seiji},
  journal = {Physical Review B},
  volume = {106},
  issue = {15},
  pages = {155153},
  numpages = {12},
  year = {2022},
  month = {Oct},
  publisher = {American Physical Society},
  doi = {10.1103/PhysRevB.106.155153},
  url = {https://link.aps.org/doi/10.1103/PhysRevB.106.155153}
}

@article{Demirplak2003,
  author  = {M. Demirplak and S. A. Rice},
  title   = {Adiabatic Population Transfer with Control Fields},
  journal = {The Journal of Physical Chemistry A},
  volume  = {107},
  number  = {46},
  pages   = {9937--9945},
  year    = {2003},
  doi     = {10.1021/jp030708a},
  url = {https://doi.org/10.1021/jp030708a}
}

@article{Demirplak2008,
  author  = {M. Demirplak and S. A. Rice},
  title   = {On the Consistency, Extremal, and Global Properties of Counterdiabatic Fields},
  journal = {The Journal of Chemical Physics},
  volume  = {129},
  pages   = {154111},
  year    = {2008},
  doi     = {10.1063/1.2992152},
  url = {https://doi.org/10.1063/1.2992152}
}

@article{Berry2009,
  author  = {M. V. Berry},
  title   = {Transitionless Quantum Driving},
  journal = {Journal of Physics A: Mathematical and Theoretical},
  volume  = {42},
  pages   = {365303},
  year    = {2009},
  doi     = {10.1088/1751-8113/42/36/365303},
  url = {https://doi.org/10.1088/1751-8113/42/36/365303}
}

@article{Chen2010,
  author  = {X. Chen and A. Ruschhaupt and S. Schmidt and A. del Campo and
             D. Gu{\'e}ry-Odelin and J. G. Muga},
  title   = {Fast Optimal Frictionless Atom Cooling in Harmonic Traps:
             Shortcuts to Adiabaticity},
  journal = {Physical Review Letters},
  volume  = {104},
  pages   = {063002},
  year    = {2010},
  doi     = {10.1103/PhysRevLett.104.063002},
  url = {https://doi.org/10.1103/PhysRevLett.104.063002}
}

@article{HammaLidar2008,
  author  = {Hamma, Alioscia and Lidar, Daniel A.},
  title   = {Adiabatic Preparation of Topological Order},
  journal = {Physical Review Letters},
  volume  = {100},
  number  = {3},
  pages   = {030502},
  year    = {2008},
  doi     = {10.1103/PhysRevLett.100.030502},
  eprint  = {quant-ph/0607145},
  archivePrefix = {arXiv},
  url = {https://doi.org/10.1103/PhysRevLett.100.030502}
}

@article{Carolan2022Impulse,
  title     = {Counterdiabatic control in the impulse regime},
  author    = {Carolan, Eoin and Kiely, Anthony and Campbell, Steve},
  journal   = {Physical Review A},
  volume    = {105},
  number    = {1},
  pages     = {012605},
  year      = {2022},
  month     = {1},
  publisher = {American Physical Society},
  doi       = {10.1103/PhysRevA.105.012605},
  url = {https://doi.org/10.1103/PhysRevA.105.012605}
}

@article{GomezCadavid2024Impulse,
  title     = {Efficient digitized counterdiabatic quantum optimization algorithm within the impulse regime for portfolio optimization},
  author    = {Gomez Cadavid, Alejandro and Montalban, Iraitz and Dalal, Archismita and Solano, Enrique and Hegade, Narendra N.},
  journal   = {Physical Review Applied},
  volume    = {22},
  number    = {5},
  pages     = {054037},
  year      = {2024},
  month     = {11},
  publisher = {American Physical Society},
  doi       = {10.1103/PhysRevApplied.22.054037},
  url = {https://doi.org/10.1103/PhysRevApplied.22.054037}
}

@incollection{Torrontegui2013,
  author    = {E. Torrontegui and S. Ib{\'a}{\~n}ez and S. Mart{\'i}nez-Garaot and
               M. Modugno and A. del Campo and D. Gu{\'e}ry-Odelin and
               A. Ruschhaupt and X. Chen and J. G. Muga},
  title     = {Shortcuts to Adiabaticity},
  booktitle = {Advances in Atomic, Molecular, and Optical Physics},
  editor    = {E. Arimondo and P. R. Berman and C. C. Lin},
  publisher = {Academic Press},
  volume    = {62},
  pages     = {117--169},
  year      = {2013},
  doi       = {10.1016/B978-0-12-408090-4.00002-5},
  url = {https://doi.org/10.1016/B978-0-12-408090-4.00002-5}
}

@article{Sung2021HighFidelity,
  author  = {Sung, Youngkyu and others},
  title   = {Realization of High-Fidelity CZ and ZZ-Free iSWAP Gates with a Tunable Coupler},
  journal = {Physical Review X},
  volume  = {11},
  pages   = {021058},
  year    = {2021},
  doi     = {10.1103/PhysRevX.11.021058},
  url = {https://doi.org/10.1103/PhysRevX.11.021058}
}

@article{Li2023SingleQubitError,
  author  = {Li, Zhiyuan and Liu, Pei and Zhao, Peng and others},
  title   = {Error per single-qubit gate below $10^{-4}$ in a superconducting qubit},
  journal = {npj Quantum Information},
  volume  = {9},
  pages   = {111},
  year    = {2023},
  doi     = {10.1038/s41534-023-00781-x},
  url = {https://doi.org/10.1038/s41534-023-00781-x}
}

@article{Xu2020TunableCoupler,
  author  = {Xu, Yuan and Chu, Ji and Yuan, Jiahao and others},
  title   = {High-fidelity, high-scalability two-qubit gate scheme for superconducting qubits},
  journal = {Physical Review Letters},
  volume  = {125},
  pages   = {240503},
  year    = {2020},
  doi     = {10.1103/PhysRevLett.125.240503},
  url = {https://doi.org/10.1103/PhysRevLett.125.240503}
}

@article{Nguyen2019Fluxonium,
  author  = {Nguyen, Long B. and Lin, Yen-Hsiang and Somoroff, Aaron and
             Mencia, Raymond and Grabon, Nicholas and Manucharyan, Vladimir E.},
  title   = {High-coherence fluxonium qubit},
  journal = {Physical Review X},
  volume  = {9},
  pages   = {041041},
  year    = {2019},
  doi     = {10.1103/PhysRevX.9.041041},
  url = {https://doi.org/10.1103/PhysRevX.9.041041}
}

@article{Verjauw2022ManufacturableQubits,
  author  = {Verjauw, J. and Acharya, R. and Van Damme, J. and others},
  title   = {Path toward manufacturable superconducting qubits with relaxation times exceeding $0.1$ ms},
  journal = {npj Quantum Information},
  volume  = {8},
  pages   = {93},
  year    = {2022},
  doi     = {10.1038/s41534-022-00600-1},
  url = {https://doi.org/10.1038/s41534-022-00600-1}
}

@article{Zhang2021FastFluxControl,
  author  = {Zhang, Helin and Chakram, Srivatsan and Roy, Tanay and others},
  title   = {Universal fast-flux control of a coherent, low-frequency qubit},
  journal = {Physical Review X},
  volume  = {11},
  pages   = {011010},
  year    = {2021},
  doi     = {10.1103/PhysRevX.11.011010},
  url = {https://doi.org/10.1103/PhysRevX.11.011010}
}

@book{LovaszPlummer1986,
  author    = {L{\'a}szl{\'o} Lov{\'a}sz and Michael D. Plummer},
  title     = {Matching Theory},
  publisher = {North-Holland},
  year      = {1986},
  volume    = {121},
  series    = {North-Holland Mathematics Studies},
  isbn      = {9780444879165},
  url       = {https://shop.elsevier.com/books/matching-theory/plummer/978-0-444-87916-5}
}

@article{Blais2021,
  title={Circuit quantum electrodynamics},
  author={Blais, Alexandre and Grimsmo, Arne L. and Girvin, Steven M. and Wallraff, Andreas},
  journal={Reviews of Modern Physics},
  volume={93},
  number={2},
  pages={025005},
  year={2021},
  publisher={American Physical Society},
  doi={10.1103/RevModPhys.93.025005},
  url = {https://doi.org/10.1103/RevModPhys.93.025005}
}

@article{An2016,
  author  = {S. An and D. Lv and A. del Campo and K. Kim},
  title   = {Shortcuts to Adiabaticity by Counterdiabatic Driving in Trapped-Ion Transport},
  journal = {Nature Communications},
  volume  = {7},
  pages   = {12999},
  year    = {2016},
  doi     = {10.1038/ncomms12999},
  url = {https://doi.org/10.1038/ncomms12999}
}

@misc{SinghYan2025GlobalGates,
  title = {Unlocking the power of global quantum gates with machine learning},
  author = {Singh, Vinit and Yan, Bin},
  year = {2025},
  eprint = {2502.02405},
  archivePrefix = {arXiv},
  primaryClass = {quant-ph},
  url = {https://arxiv.org/abs/2502.02405}
}

@Article{Visuri2026,
author={Visuri, Anne-Maria
and Gomez Cadavid, Alejandro
and Bhargava, Balaganchi A.
and Romero, Sebasti{\'a}n V.
and Grabarits, Andr{\'a}s
and Chandarana, Pranav
and Solano, Enrique
and del Campo, Adolfo
and Hegade, Narendra N.},
title={Digitized counterdiabatic quantum critical dynamics},
journal={npj Quantum Information},
year={2026},
month={Mar},
day={13},
volume={12},
number={1},
pages={47},
issn={2056-6387},
doi={10.1038/s41534-026-01208-z},
url={https://doi.org/10.1038/s41534-026-01208-z}
}

\end{document}